\documentclass[twocolumn]{aastex62}
\usepackage{lineno}
\hypersetup{linkcolor=red,citecolor=blue,filecolor=cyan,urlcolor=magenta}

\newcommand{\kms}{\mbox{km\,s$^{-1}$}}
\newcommand{\htwo}{\mbox{$\rm H_2$}}
\newcommand{\methanol}{\mbox{$\rm CH_3OH$}}
\newcommand{\ntwodp}{\mbox{$\rm N_2D^+$}}
\newcommand{\ntwohp}{\mbox{$\rm N_2H^+$}}
\newcommand{\ceighteeno}{\mbox{$\rm C^{18}O$}}
\newcommand{\thirteenco}{\mbox{$\rm ^{13}CO$}}
\newcommand{\dcop}{\mbox{$\rm DCO^+$}}

\newcommand{\dcn}{\mbox{$\rm DCN$}}
\newcommand{\ammonia}{\mbox{$\rm NH_3$}}
 
\newcommand{\mjypbm}{\mbox{mJy\,beam$^{-1}$}}
\newcommand{\jypbm}{\mbox{Jy\,beam$^{-1}$}}
\newcommand{\Dfrac}{\mbox{$D_{\rm frac}$}}
\newcommand{\Tdust}{\mbox{$T_{\rm dust}$}}
\newcommand{\fratio}{\mbox{$f_{\rm 1.05mm}/f_{\rm 1.30mm}$}}
\newcommand{\msun}{\mbox{\,$M_\odot$}}

\usepackage{amsmath}

\graphicspath{{./}{figures/}}

\received{}
\revised{}
\accepted{}

\shorttitle{Star Formation in a Strongly Magnetized Cloud}
\shortauthors{Cheng et al.}

\begin{document}

\title{Star Formation in a Strongly Magnetized Cloud}

\correspondingauthor{Yu Cheng}
\email{ycheng.astro@gmail.com}

\author[0000-0002-0786-7307]{Yu Cheng}
\affil{Dept. of Astronomy, University of Virginia, Charlottesville, Virginia 22904, USA}

\author{Jonathan C. Tan}
\affil{Dept. of Astronomy, University of Virginia, Charlottesville, Virginia 22904, USA}
\affil{Dept. of Space, Earth \& Environment, Chalmers University of Technology, Gothenburg, Sweden}

\author{Paola Caselli}
\affiliation{Max-Planck-Institute for Extraterrestrial Physics (MPE), Giessenbachstr. 1, D-85748 Garching, Germany}

\author{Laura Fissel}
\affil{Engineering Physics and Astronomy, Queen's University, Kingston, ON K7L 3N6, Canada}

\author{H{\'e}ctor G. Arce}
\affiliation{Department of Astronomy, Yale University, New Haven, CT 06511, USA} 

\author{Francesco Fontani}
\affil{INAF-Osservatorio Astrofisico di Arcetri, Largo E. Fermi 5, I-50125, Florence, Italy}
\affil{Centre for Astrochemical Studies, Max-Planck-Institute for Extraterrestrial Physics, Giessenbachstrasse 1, D-85748 Garching, Germany}

\author{Matthew D. Goodson}
\affiliation{Dept. of Physics and Astronomy, University of North Carolina at Chapel Hill, Chapel Hill, NC 27599-3255, USA}

\author{Mengyao Liu}
\affil{Dept. of Astronomy, University of Virginia, Charlottesville, Virginia 22904, USA}

\author{Nicholas Galitzki}
\affiliation{University of California San Diego, La Jolla, CA, United States}

\begin{abstract}
We study star formation in the Center Ridge 1 (CR1) clump in the Vela C giant molecular cloud, selected as a high column density region that shows the lowest level of dust continuum polarization angle dispersion, likely indicating that the magnetic field is relatively strong. We observe the source with the ALMA 7m-array at 1.05~mm and 1.3~mm wavelengths, which enable measurements of dust temperature, core mass and astrochemical deuteration. A relatively modest number of eleven dense cores are identified via their dust continuum emission, with masses spanning from 0.17 to 6.7~{$M_\odot$}. Overall CR1 has a relatively low compact dense gas fraction compared with other typical clouds with similar column densities, which may be a result of the strong magnetic field and/or the very early evolutionary stage of this region.
The deuteration ratios, \Dfrac{}, of the cores, measured with \ntwohp(3-2) and \ntwodp(3-2) lines, span from 0.011 to 0.85, with the latter being one of the highest values yet detected. The level of deuteration appears to decrease with evolution from prestellar to protostellar phase. A linear filament, running approximately parallel with the large scale magnetic field orientation, is seen connecting the two most massive cores, each having CO bipolar outflows aligned orthogonally to the filament. The filament contains the most deuterated core, likely to be prestellar and located midway between the protostars. The observations permit measurement of the full deuteration structure of the filament along its length, which we present. We also discuss the kinematics and dynamics of this structure, as well as of the dense core population.
\end{abstract}


\section{Introduction}\label{sec:intro}

Star formation is a complicated process with many open questions, including what sets its rate, overall efficiency, and resulting mass distribution of stars, i.e., the stellar initial mass function (IMF). To help answer these questions, detailed studies of star-forming regions that can resolve individual self-gravitating cores are needed and these regions should span as wide a range of environmental conditions as possible in order to explore potential effects of these conditions. With this goal in mind, we present here a study of a dense star-forming clump in the Vela C giant molecular cloud (GMC) that has been selected to have a low angular dispersion in sub-mm polarization position angles, which likely indicates that it has relatively strong magnetic fields.

The Vela molecular cloud complex is one of the nearest giant molecular cloud complexes in the Galactic disk \citep{Murphy91}. It is composed of four molecular clouds, of which Vela C is the most massive and the host of the youngest stellar population \citep{Yamaguchi99}. Vela C is known to harbor low, intermediate and high-mass star formation \citep{Massi03,Netterfield09} and hence is an ideal laboratory to study different modes of star formation. When contoured at $A_V$ = 7 mag, the Vela C cloud appears to segregate into five distinct sub regions (North, Centre-Ridge, Centre-Nest, South-Ridge, and South-Nest), each with distinct morphological characteristics \citep{Hill11}. In the Centre-Ridge sub region there is a compact HII region, RCW36, which is adjacent to a very prominent dust ridge that hosts the majority of dense cores in the cloud \citep{Hill11}. Owing to its proximity, i.e., at a distance of $933\pm94$~pc \citep{Fissel19}, 
Vela C has been an important target for magnetic field mapping studies through sub-mm polarimetry and near-infrared stellar polarimetry \citep{Fissel16,Kusune16,Santos17}. In particular, the relative orientation between gas column density filamentary 
structures and the magnetic field changes progressively with increasing gas column density, from mostly parallel or having no preferred orientation at low column densities to mostly perpendicular at the highest column densities \citep{Soler17,Fissel19}. This suggests that the magnetic field is strong enough to have influenced the formation of the dense gas structures within Vela C.

\begin{deluxetable*}{lccccc}
\tabletypesize{\scriptsize}
\caption{Observed transitions}
\label{table:transitions}
\tablehead{
\colhead{molecular transition} & \colhead{frequency$^{a}$} & \colhead{$\rm E_u/k$} & \colhead{HPBW} & \colhead{$\triangle v$} & \colhead{sensitivity}\\
 & (GHz) & (K) & (\arcsec) & (\kms) & (\jypbm~ per channel)
}
\startdata
\ntwodp(3-2)   & 231.321912  & 22.2  & 7.07\arcsec$\times$4.44\arcsec & 0.046 & 0.20\\
\thirteenco(2-1) & 220.398684&  15.9 &7.63\arcsec$\times$4.57\arcsec & 0.096 & 0.20\\
\ceighteeno(2-1) & 219.560354&  15.8 & 7.65\arcsec$\times$4.61\arcsec& 0.096 & 0.16\\
\dcn(3-2)   & 217.238538     &  20.9  & 7.48\arcsec$\times$4.84\arcsec & 0.195 & 0.10\\
SiO(5-4) & 217.104980    &  31.3 & 7.49\arcsec$\times$4.84\arcsec& 0.195 &0.09\\
\methanol($\rm 5_{1,4}-4_{2,2}$)&216.945521& 55.9 &7.50\arcsec$\times$4.84\arcsec &0.196 &0.09\\
\dcop(3-2) & 216.112580      &    20.7 &7.50\arcsec$\times$4.86\arcsec &0.196 &0.09 \\
\ntwohp(3-2) & 279.511832    &  26.8  &5.88\arcsec$\times$3.60\arcsec & 0.038 & 0.30\\
\dcn(4-3) & 289.644907       &  34.8   & 5.75\arcsec$\times$3.51\arcsec& 0.073 & 0.20\\
\dcop(4-3) & 288.143858      &    34.6 & 5.71\arcsec$\times$3.51\arcsec& 0.073 &0.30
\enddata
\tablenotetext{a}{Line frequencies from Cologne Database for Molecular Spectroscopy (CDMS; http://www.astro.uni-koeln.de/cdms/catalog) \citep{Muller05}. For \ntwohp(3-2) and \ntwodp(3-2) we list the frequency of the hyperfine component with the largest $\rm A_{ul}$ emission coefficient in \citet{Pagani09}. }
\end{deluxetable*}

The ongoing star formation in Vela C has been investigated in several studies via far-infrared (FIR) to mm continuum imaging \citep[e.g.,][]{Netterfield09,Giannini12,Massi19}. \citet{Giannini12} identified 268 dense cores with {\it Herschel} FIR data. \citet{Massi19} found 549 cores based on sub-mm continuum mapping using APEX and derived a prestellar core mass function (CMF) that has a similar shape as the stellar IMF at the high mass end. However, these observations are limited by their relatively low spatial resolution, i.e., $\gtrsim20\arcsec$ (0.09~pc), which is unable to resolve down to the scale of dense cores (i.e., a few $\times$~0.01~ pc) relevant to the formation of individual stars or small-$N$ multiple systems. 

\begin{figure*}[ht!]
\epsscale{1.0}\plotone{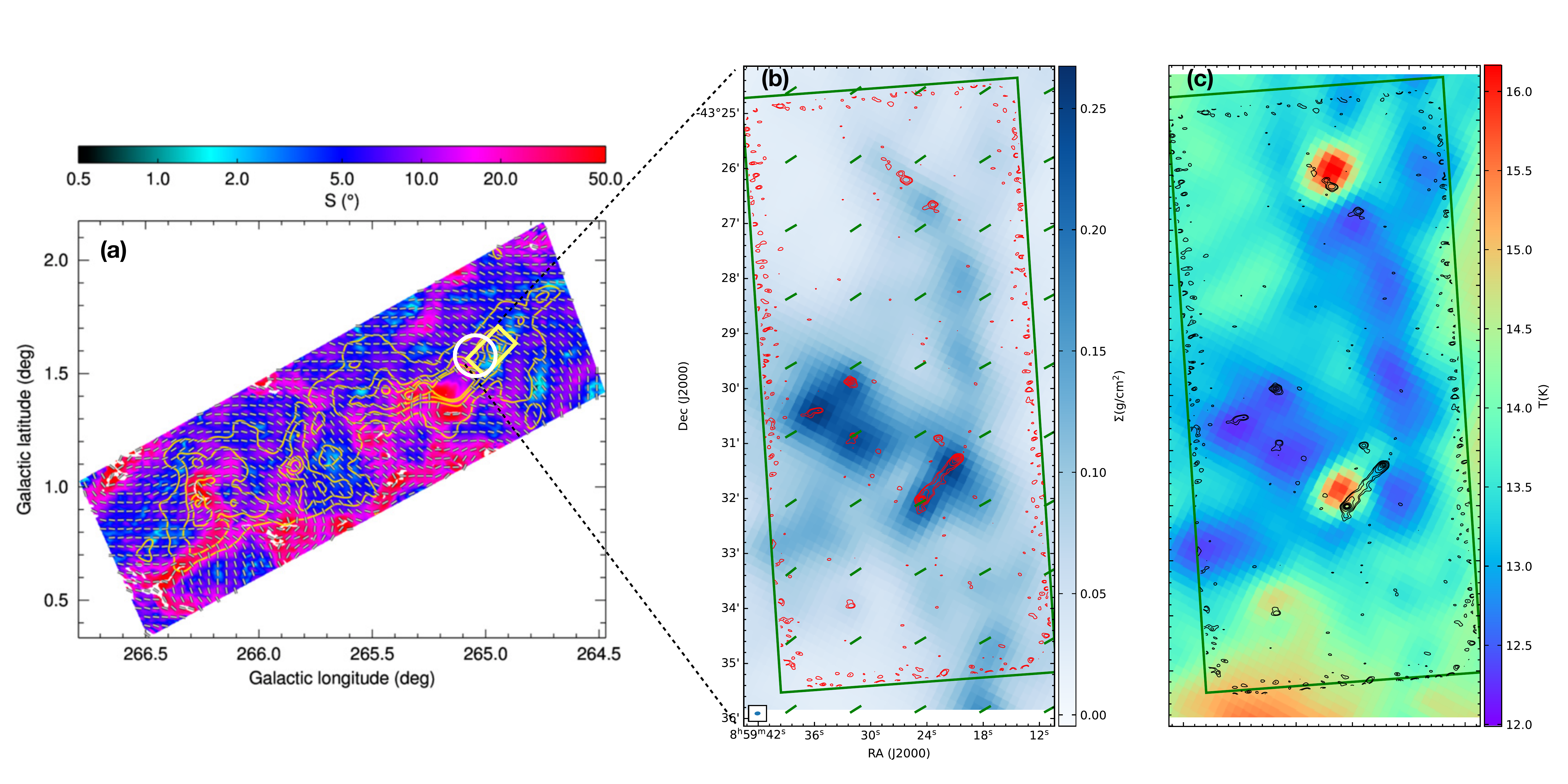}
\caption{(a) Figure 6 from \citet{Fissel16}. BLASTPol map of the dispersion in the polarization-angle in degrees on 0.5 pc scales, shown in colorscale. Line segments show the orientation of the magnetic field as projected on the plane of the sky ($\Phi$), derived from the BLASTPol 500 $\mu$m data. The $\Phi$ measurements are shown approximately every 2.5\arcmin. Contours indicate 500 $\mu$m I intensity levels of 46, 94, 142, and 190 MJy sr$^{-1}$. The yellow box indicates the region mapped by ALMA in this study, which is selected based on its appearance as a local minimum on the polarization-angle dispersion map. The position of the \#5 \ceighteeno{} clump in \citet{Yamaguchi99} is indicated with a white circle with a radius of 4\arcmin. (b) Mass surface density map derived with the {\it Herschel} data shown in color scale. The red contours indicate the ALMA 1.3 mm continuum map. The contour levels are $\sigma$ \  $\times$ (4, 6, 10, 20, 40, 80), with 1$\sigma$ = 1.3 \mjypbm. The direction of the POS magnetic field in panel (a) is shown in green line segments. (c) Temperature map derived with the {\it Herschel} data shown in color scale. The black contours show the ALMA 1.3 mm continuum map.
}
\label{fig:overview}
\end{figure*}

In this paper we present an ALMA 7m-array study in both Band 6 and Band 7 towards a dense clump in the Center Ridge of Vela C (referred as CR1 clump hereafter) and the observations achieve $\sim$5\arcsec\ resolution for various molecular species (see \autoref{table:transitions}). 
The CR1 clump is located to the north of a hot pocket of gas (RCW 36) around the OB cluster, but appears to not yet be impacted by it \citep{Hill11}. The CR1 clump has been selected for this study because it appears to be strongly magnetized as evidenced by having a local minimum of angular dispersion in sub-mm polarization position angles, as shown in \autoref{fig:overview} \citep[see also Figure 6 in][]{Fissel16}. Thus, the main goal of this paper is to study the dense core population leading to star formation in this example of a strongly magnetized environment. The CR1 clump is close to the \#5 \ceighteeno{} clump identified in \citet{Yamaguchi99} (see also \autoref{fig:overview}), for which \citet{Kusune16} estimated a plane-of-the-sky (POS) magnetic field strength of 120 $\mu$G based on near-IR stellar polarimetry. According to the Chandrasekhar-Fermi method \citep{Chandrasekhar53}, the POS magnetic field strength $\rm B_{pos}$ can be expressed as

\begin{equation}
    B_{\rm pos} = Q \sqrt{4\pi \rho} \frac{\sigma_v}{\sigma_\theta}
\end{equation}
where $\rho$ is the mean density of the cloud, $\sigma_v$ is the line-of-sight velocity dispersion, $\sigma_{\theta}$ is the dispersion of the polarization position angles, and $Q\sim$ 0.5 is a correction factor for $\sigma_{\theta} \lesssim 25^{\circ}$ \citep{Ostriker01}. In \citet{Kusune16} the angular dispersion of polarization angles in the \#5 \ceighteeno{} clump is estimated to be 18$^{\circ}$. However, it is difficult to probe the magnetic field structures in high extinction regions with near-IR polarimetry and most polarization vectors are from the relative diffuse part of the cloud. The angular dispersion of our mapped region (or \#5 \ceighteeno{} clump) appears much lower in the BLASTPol survey, i.e., $\sim 2 ^{\circ}$ (see also \autoref{fig:overview}), leading to a higher $\rm B_{pos}$ estimation of $\sim$ 1 mG. Note that it is likely that the small scale magnetic field variation is not resolved in the BLASTPol survey (reso. $\sim$ 2.5\arcmin), so future high resolution dust polarization observations are required to clarify the field strength in this region. Nevertheless, the selected region is likely to have a relatively strong magnetic field compared to surrounding regions in Vela C.

Our Band 6 spectral set-up and analysis methods are similar to our previous studies of the G286 protocluster \citep{Cheng20}
and Infrared Dark Clouds (IRDCs) \citep[e.g.,][]{Tan13,Kong17}, which have a goal of studying cores via their mm dust continuum emission and via emission lines from dense gas tracers, especially \ntwodp(3-2). The Band 7 spectral set-up is designed to obtain a sub-mm dust continuum measurement, as well as observation of \ntwohp(3-2) that allows an accurate estimate of the level of deuteration of \ntwohp, which is expected to be boosted in cold, dense conditions and thus may be a useful evolutionary indicator of prestellar and early stage protostellar cores.


This paper is structured as follows: the observations and results are presented in \autoref{sec:observations} and \autoref{sec:results}, respectively. We further discuss our results in \autoref{sec:discussion}, and  present our conclusions in \autoref{sec:conclusion}.

\begin{figure*}[ht!]
\epsscale{1.1}\plotone{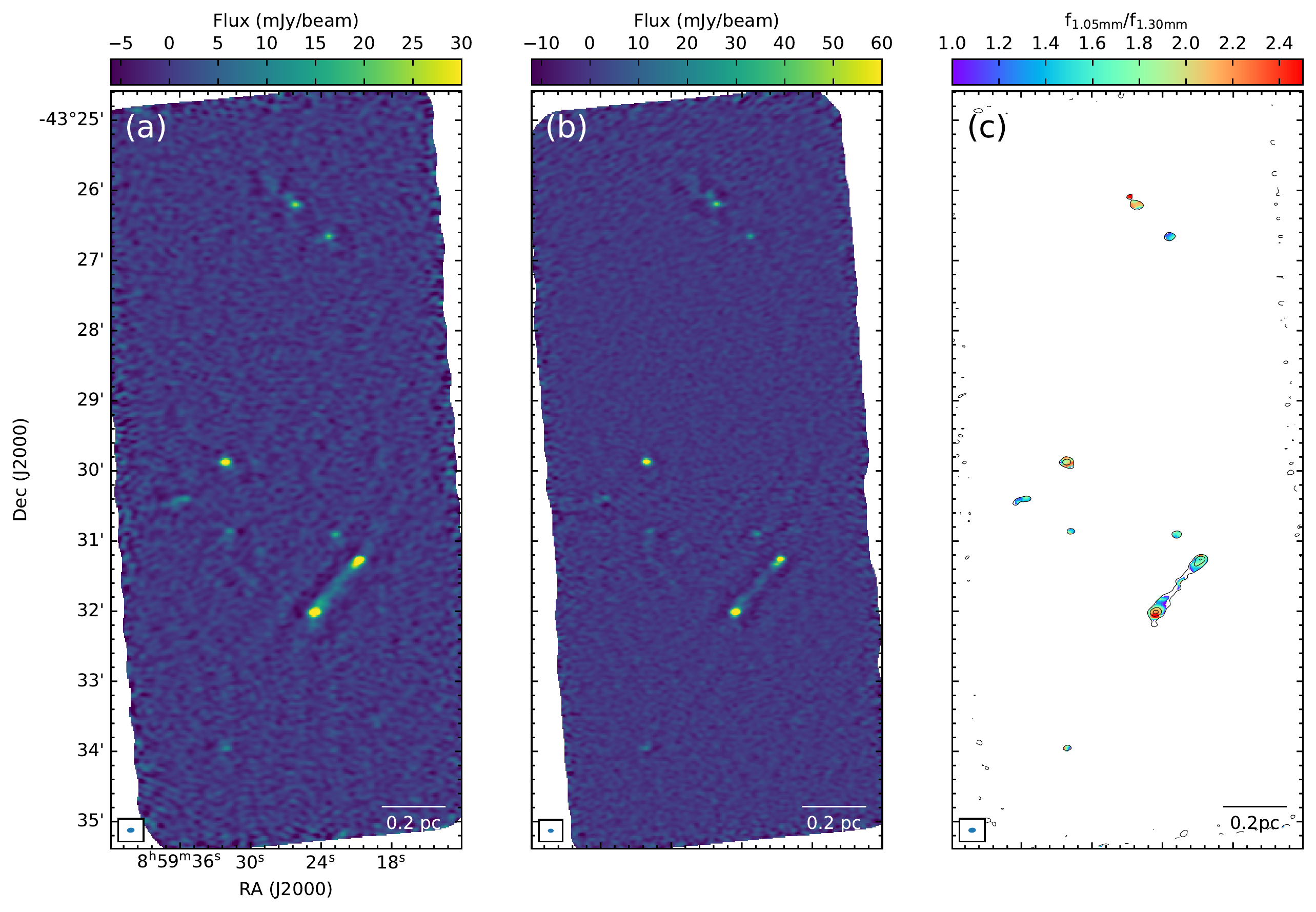}
\caption{(a) ALMA Band 6 (1.3 mm) continuum image of Vela C CR1. 
{(b)} ALMA Band 7 (1.05 mm) continuum image of Vela C CR1.  
{(c)} Map of flux ratio $\rm f_{1.05mm}/f_{1.30mm}$. Only the regions with flux above 3$\sigma$ in both bands are shown. In deriving this map we found a systematic positional offset $\sim$0\farcs{3} between Band~6 and Band~7 maps, possibly due to imperfect phase calibration. This offset has been corrected in this map.
}
\label{fig:cont}
\end{figure*} 

\section{Observations}
\label{sec:observations}
\subsection{ALMA observations}
The observations were conducted with the ALMA 7m-array in Bands 6 and 7 in Cycle 6 (Project ID
2018.1.00227.S, PI: J. C. Tan), during a period from March to
April 2019. The entire field (10\arcmin$\times$4.5\arcmin) was divided into four 
strips, each about 150\arcsec\ wide and 270\arcsec\ long. 

For the Band 6 observation we set the central frequency of the
correlator sidebands to be the rest frequency of the
$\rm{N_2D}^+$(3-2) line for SPW0 with a velocity
resolution of 0.046 \kms. The second baseband
SPW1 was set to $231.00\:$ GHz, i.e., 1.30~mm, to observe the continuum
with a total bandwidth of $1.875\:$ GHz, which also covers CO(2-1) with a
velocity resolution of 1.46~\kms. 
SPW2 was split to cover $\rm{^{13}CO}$(2-1) and 
$\rm{C^{18}O}$(2-1) line, both with a velocity 
resolution of 0.096~\kms. 
The frequency coverage for SPW3
ranged from 215.85 to $217.54\:$GHz to observe DCN(3-2), DCO$^+$(3-2),
SiO(5-4) and $\rm{CH_3OH}(5_{1,4}-4_{2,2})$. 

For Band 7 we set the central frequency to be the rest frequency of the
$\rm{N_2H}^+$(3-2) line for SPW0 with a velocity resolution 
of 0.038~\kms. The central frequencies of SPW1 and SPW2 were set to 278.88~GHz
and 291.10~GHz, respectively, and each band had a bandwidth of $1.875\:$GHz to
observe continuum emission. SPW3 was split equally to
observe two lines, i.e., DCN(4-3), DCO$^+$(4-3), with 
58.59 MHz (61 km/s) bandwidth and resolution of 0.073~km/s.

The raw data were calibrated with the data reduction pipeline using
{\it{Casa}} 5.4.0. The continuum visibility data were constructed with
all line-free channels. We performed imaging with the {\it tclean} task in
{\it Casa} and during cleaning we combined data for all four strips to
generate a final mosaic map. The 7m-array data were
imaged using a Briggs weighting scheme with a robust parameter of 0.5,
which yields a resolution of $7.00\arcsec\times4.29\arcsec\:$ for Band 6,
and $5.92\arcsec\times3.47\arcsec\:$ for Band 7. 
The $1\sigma$ noise levels in the continuum image are 1.3 \mjypbm~ and
1.8 \mjypbm~ for Band 6 and Band 7, respectively. The resolutions and 
sensitivities for spectral lines are summarized in \autoref{table:transitions}.


\subsection{Auxiliary data}

We have retrieved archival data to provide auxiliary information at infrared wavelengths. The 3.5 and 4.5 $\mu$m maps are from the Spitzer Heritage Archive hosted in the NASA/IPAC Infrared Science Archive. For 12 and 22~$\mu$m we use Wide-field Infrared Survey Explorer {\it WISE} archival data. Continuum images in the wavelengths of {\it Herschel PACS} (70 and 160 $\mu$m) and {\it SPIRE} (250, 350, and 500 $\mu$m) were obtained from the Herschel Science Archive. For this, Vela~C was observed on 2010, May 18, as part of the HOBYS (Herschel imaging survey of OB young stellar objects, \citet{Motte10}) guaranteed time key program. 

We also obtained the total hydrogen column density $N_{\rm H}$ (in units of hydrogen
nuclei per cm$^{-2}$) and temperature map (see \autoref{fig:overview}), which were first presented in Section 5 of \citet{Fissel16}. These maps are based on dust spectral fits to four far-IR/sub-mm dust emission maps: {\it Herschel}-SPIRE maps at 250, 350, and 500 $\mu$m; and a 
{\it Herschel}-PACS map at 160 $\mu$m. These maps have the same spatial resolutions as the 500 $\mu$m map, i.e., 35.2\arcsec.

\begin{figure}[ht!]
\epsscale{1.0}\plotone{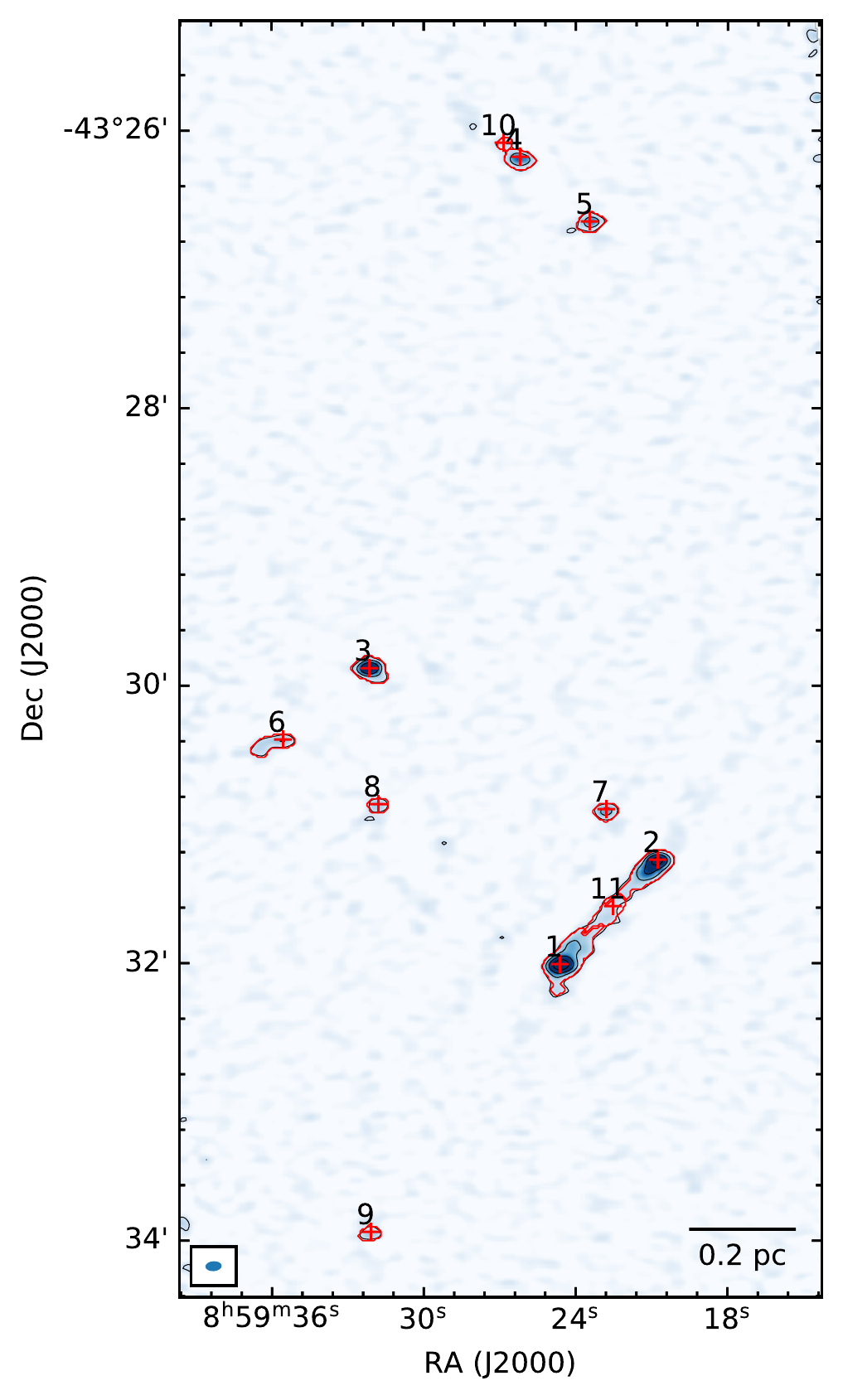}
\caption{Cores identified with {\it dendrogram} overlaid on the 1.3 mm continuum. The red crosses indicate the peak positions, while the red contours indicate the boundaries returned by {\it dendrogram}. Note that CR1c11 was identified via \ntwodp{} moment 0 map (see text).
}
\label{fig:dendro}
\end{figure} 

\begin{figure}[ht!]
\epsscale{1.0}\plotone{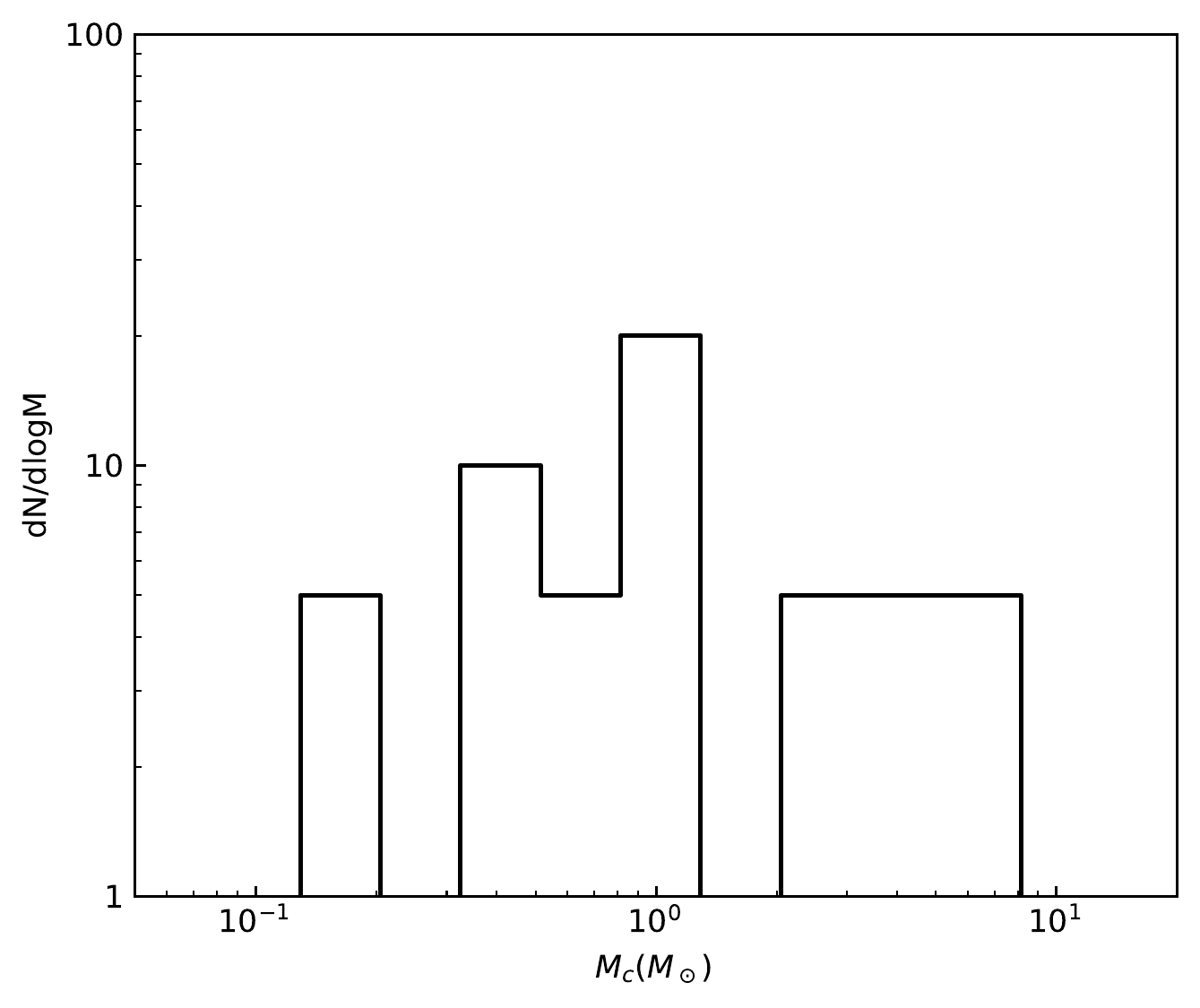}
\caption{The mass distribution of cores detected in the Vela C CR1 region.}
\label{fig:cmf}
\end{figure} 

\section{Results}
\label{sec:results}
\subsection{Continuum}
\label{sec:cont}

\begin{deluxetable*}{cccccccccccc}
\tabletypesize{\scriptsize}
\caption{Core properties}
\label{table:core_property}
\tablehead{
\colhead{Core} & \colhead{R.A.}& \colhead{Dec.}  & \colhead{$M_c$} & \colhead{Area} & \colhead{$R_c(")$} & \colhead{$R_c$} &\colhead{$\Sigma_c $} & \colhead{$n_{{\rm H},c}$ }& \colhead{$\frac{f_{\rm 1.05mm}}{f_{\rm 1.30mm}}$} & \colhead{$T_c$\tablenotemark{a}} & \colhead{$\rm \alpha_{vir}$\tablenotemark{b}}  \\
\colhead{} & \colhead{($^{\circ}$)} & \colhead{($^{\circ}$)} &  \colhead{($M_\odot$)} & \colhead{arcsec$^2$}&\colhead{arcsec}& \colhead{(0.01pc)} &\colhead{(g cm$^{-2}$)} & \colhead{10$^5$cm$^{-3}$}&   &  \colhead{(K)  } & \colhead{}
}
\startdata
1 & 134.85254 & -43.53361 & 6.69 & 301 & 9.47 &4.27 & 0.228 & 6.00 & 1.83 $\pm$ 0.14 & $\rm7.8 ^{+3.6}_{-1.9}$ & 1.30\\
2 & 134.83645 & -43.52111 & 4.86 & 206 & 7.83 & 3.53 & 0.242 & 7.70 & 1.59 $\pm$ 0.13 & $\rm5.1 ^{+1.2}_{-0.8}$ & 1.01\\
3 & 134.88394 & -43.49805 & 2.52 & 117 & 5.90 &2.66 & 0.222 & 9.35 & 1.90 $\pm$ 0.16 & $\rm9.4 ^{+7.6}_{-2.8}$ & 0.98\\
4 & 134.85906 & -43.43667 & 1.19 & 76 &4.77 &  2.15 & 0.161 & 8.41 & 2.02 $\pm$ 0.21 & $\rm13.6 ^{+116.6}_{-6.1}$ & 1.58\\
5 & 134.84758 & -43.44444 & 0.88 & 68 &4.50 & 2.03 & 0.133 & 7.35 & 1.37 $\pm$ 0.19 & $\rm3.8 ^{+1.0}_{-0.7}$ & 1.47\\
6 & 134.89812 & -43.50666 & 0.88 & 92 &5.23 & 2.36 & 0.098 & 4.66 & 1.36 $\pm$ 0.21 & $\rm3.8 ^{+1.2}_{-0.8}$ & 3.00\\
7 & 134.84488 & -43.51500 & 0.61 & 52 &3.95 & 1.78 & 0.120 & 7.59 & 1.59 $\pm$ 0.25 & $\rm5.1 ^{+2.9}_{-1.4}$ &\\ 
8 & 134.88242 & -43.51444 & 0.35 & 37 &3.33 & 1.50 & 0.097 & 7.30 & 1.63 $\pm$ 0.35 & $\rm5.5 ^{+6.3}_{-2.0}$ &\\ 
9 & 134.88358 & -43.56583 & 0.34 & 37 &3.33 & 1.50 & 0.094 & 7.04 & 1.63 $\pm$ 0.36 & $\rm5.4 ^{+6.7}_{-2.0}$ &\\ 
10 & 134.86173 & -43.43500 & 0.17 & 20 & 2.44& 1.10 & 0.087 & 8.88 & 2.48 $\pm$ 0.68 & $>$7.4 &\\ 
11 & 134.84373 & -43.52667 & 0.88 & 110 &7.72 & 2.58 & 0.082 & 3.56 & 1.28 $\pm$ 0.22 & $\rm3.5 ^{+1.0}_{-0.7}$ & 4.84
\enddata
\tablenotetext{a}{Estimated from ratio of $f_{\rm 1.05mm}/f_{\rm 1.30mm}$ assuming optically thin thermal emission from dust and dust opacities of the moderately coagulated thin ice mantle model of \citet{Ossenkopf94}.}
\tablenotetext{b}{For each core the virial parameter is derived with a deconvolved core radius, and velocity dispersions combining measurements with different tracers, i.e., the same as panel (d) in \autoref{fig:vir}. }
\end{deluxetable*}

\autoref{fig:cont} illustrates the Band 6 (1.3~mm) and Band 7 (1.05~mm) continuum of the Vela C CR1 clump. Overall there are about 10 clearly visible cores sparsely distributed over the field.
The detections at 1.05~mm are similar to those at 1.3~mm. The two brightest cores are located in the southern part of the field, with a linear filament or ``bridging feature'' connecting them. This bridge is about 0.27~pc long and appears more prominent at 1.3~mm. As shown in \autoref{fig:overview}, the orientation of this bridging feature is close to the POS direction of the magnetic field derived in the BLASTPol survey, with an offset of $\sim$ 18 $^{\circ}$.  

We used the {\it dendrogram} algorithm \citep{Rosolowsky08} implemented with {\it astrodendro} to carry out an automated, systematic search for cores 
in the continuum images following the method used in \citet{Cheng18} and \citet{Liu18}.
We defined the identified leaves (the base element in the hierarchy of dendrogram that has no further sub-structure) as cores. We set the minimum flux density threshold to $4\sigma$, the minimum significance for structures to $1\sigma$, and the minimum area to half the size of the synthesized beam. We tried {\it dendrogram} identification on the continuum maps of both bands and found almost equivalent results. Hereafter we define the positions and boundaries of cores based on the 1.3~mm data, which have slightly better signal to noise ratios, as shown in \autoref{fig:dendro}. The cores are named as CR1c1, CR1c2, etc., with the numbering order from highest to lowest integrated flux. There is an additional core (CR1c11) that is located at the bridging feature and not identified as a core from the 1.3~mm data, but it does appear as an independent condensation in 1.05~mm continuum, and moment 0 maps of some lines like \ntwodp{}(3-2) and \dcop{}(3-2). So we also include CR1c11 in our sample and adopt a core boundary defined using the \ntwodp{} moment 0 map (by running {\it dendrogram} with the same set up). Then the regions of CR1c1 and CR1c2 that overlap with CR1c11 are excluded from the definition of CR1c1 and CR1c2 when deriving their properties.
 
We then estimated the masses of cores assuming the 1.3~mm emission comes from optically thin thermal dust emission with a uniform temperature of 15~K following the methods and assumptions used in the study of \citet{Cheng18}, with the only difference being that this previous study adopted a fiducial temperature of 20~K. Our reason to choose a slightly lower temperature is the availability in Vela C of a relatively high resolution temperature map (though not high enough to resolve individual cores themselves) that indicates temperatures closer to 15~K. The estimated masses range from 0.17 to 6.7~$M_\odot$. If temperatures of 10~K or 20~K were to be adopted, then the mass estimates would differ by factors of 1.85 and 0.677, respectively. In \autoref{fig:cmf} we plot the CMF of the detected sample. Given the small numbers of detected cores, it is difficult to make meaningful comparison with the CMFs of other regions. 

The core radii are evaluated as $R_c=\sqrt{A/\pi}$, where $A$ is the projected area of each core returned by the {\it dendrogram} algorithm. The median radius is 0.016~pc (i.e., 3300 au), similar to the spatial resolution of 5100~au that is achieved by the 
$\sim$ 5.5\arcsec\ angular resolution observations. This indicates that most cores are not well resolved. Given masses and radii, the mass surface densities and volume number densities of the cores can be estimated. These properties are summarized in \autoref{table:core_property}. 

\begin{figure}[ht!]
\epsscale{1.2}\plotone{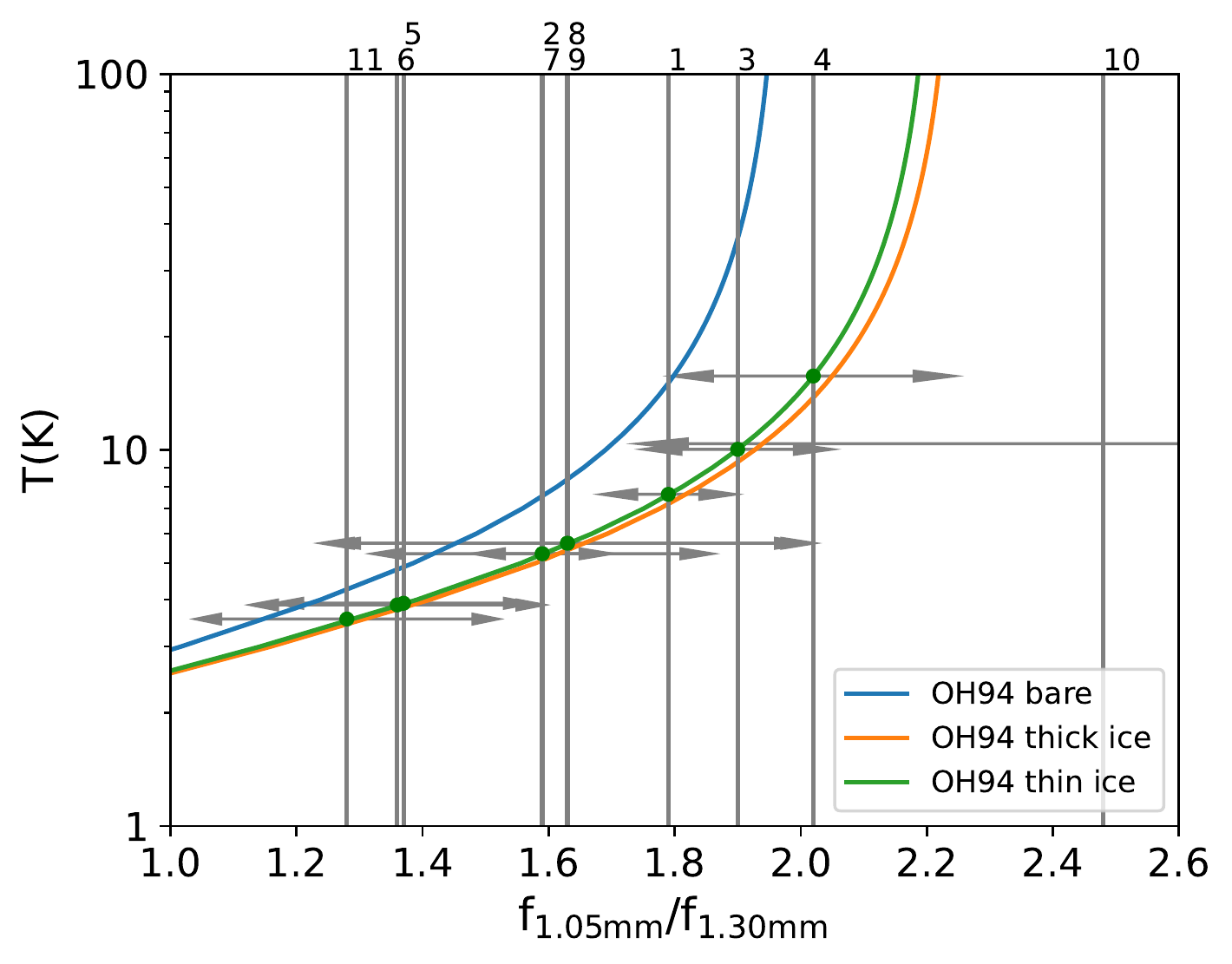}
\caption{The predicted relation between $f_{\rm 1.05mm}/f_{\rm 1.30mm}$ and temperature for three dust models from \citet{Ossenkopf94}. The measured ratios for the dense cores of Vela C CR1 are shown by the vertical lines, with uncertainties shown as two headed arrows at the location where the vertical line crosses the fiducial OH94 thin ice mantle model (or at 10~K level for CR1c10). The indices of cores as in \autoref{table:core_property} are labeled on top of the plot. 
}
\label{fig:ratio}
\end{figure}

In \autoref{fig:cont}(c) we present the map of 1.05~mm/1.3~mm flux ratio for positions with fluxes greater than $3\sigma$ in both bands (after convolving 1.05~mm data to the angular resolution of the 1.3~mm map). This ratio ranges from 1.0 to 2.5 over the map. We use this ratio to give more constraints on the dust temperature. To do this, we compare the observed ratio \fratio{} with that predicted from models of optically thin thermal dust emission, i.e.,
\begin{equation}
   \frac{f_{\nu_1}}{f_{\nu_2}} = \frac{B_{\nu_1}(T_{\rm dust})}{B_{\nu_2}(T_{\rm dust})}\cdot \frac{\kappa_{\nu_1}}{\kappa_{\nu_2}} =  \frac{B_{\nu_1}(T_{\rm dust})}{B_{\nu_2}(T_{\rm dust})}\cdot \left(\frac{\nu_1}{\nu_2}\right)^\beta
\end{equation}
where $f_\nu$ is the dust emission flux at frequency $\nu$, $B_\nu(T_{\rm dust})$ is the Planck function with dust temperature $T_{\rm dust}$, $\kappa_\nu$ is the dust opacity and $\beta$ is the dust opacity index. For fiducial dust opacity we adopt the same model that we have used for our mass estimates, i.e., the thin ice mantle model of \citet{Ossenkopf94} with $10^5$ years of coagulation at a density of $n_{\rm H}$ = $\rm 10^6 cm^{-3}$.  At sub-mm wavelengths, this model exhibits $\kappa_\nu$  = $\rm 0.1(\nu/1000\: GHz)^{\beta} cm^2 g^{-1}$ with $\beta\simeq1.8$. As shown in \autoref{fig:ratio}, \fratio{} increases from 1.5 at $T_{\rm dust}$ = 5 K to about 2.1 at $T_{\rm dust}$ = 20~K, and grows asymptotically to 2.2 at higher temperatures. For comparison, we also present the predicted \fratio-\Tdust{} relation for the equivalent bare grain and thick ice mantle models of \citet{Ossenkopf94}. We see that for the models with ice mantles (thin/thick), which are expected to be the most relevant for prestellar and early stage protostellar cores, the choice of dust opacity model does not strongly affect the derived \Tdust{} for a given flux ratio. More generally, our derived \Tdust{} estimates are valid for dust opacity models that have a spectral index $\beta$ close to 1.8 in the millimeter wavelength regime.

Given the observed values of \fratio{}, we estimate $T_{\rm dust}$ by looking for the corresponding values on the predicted relation, as shown in \autoref{fig:ratio}. The uncertainties in \fratio{} are also transferred into the uncertainties in $T_{\rm dust}$. \autoref{table:core_property} lists these derived temperatures. The fluxes of cores are measured by integrating over the region defined by {\it dendrogram} and for flux uncertainties we consider both the root-mean-square error and a flux calibration uncertainty of about 5\%, and combine them in quadrature. 

The measured \Tdust{} values range from 3.5~K to 13.6~K. For CR1c10 the \fratio{} is 2.48 $\pm$ 0.68, leading to an unrealistic \Tdust{} of $957.0_{-949.6}^{+\infty}$ K, so we only conservatively list the lower limit of 7.4~K. 
In general the derived core temperatures appear to be relatively low compared to canonical estimates of temperatures in molecular clouds, i.e., typically found to be in the range $\sim10-20$~K. On the larger scales probed by the {\it Herschel} sub-mm observations (see \autoref{fig:overview}), the CR1 clump is estimated to have dust temperatures of $\sim$ 12-16~K. Still, we note that the centers of some prestellar cores have been measured to have temperatures as low as about 6~K from $\rm NH_3$ observations \citep{Crapsi07}.
We further note that there are several potential sources of systematic uncertainties in the temperature estimation from \fratio{}. The effects of choice of dust model have already been described in \autoref{fig:ratio}. In addition, since the core boundaries are defined based on the 1.3~mm data, we expect that the estimated flux ratio \fratio{} and correspondingly \Tdust{} could be systematically underestimated. Differences in recovered flux fractions could also introduce systematic uncertainties, with a smaller flux recovery fraction generally expected at 1.05~mm. 
Another potential source of uncertainty is if the cores (or part of the cores) become optically thick, which would occur first at 1.05~mm. This would tend to lower the flux received at 1.05~mm, again causing an underestimation of $T_{\rm dust}$. For example, if a core is moderately optical thick at 1.05~mm with $\tau_{\rm 1.05mm} = 1$, then the resulting \fratio{} will be $\sim$13\% lower than the case assuming optical thin. However, most cores in our sample should be optically thin judging from the observed low brightness temperatures $T_B$. 
In Band 7 the median $T_B$ seen at the continuum peaks of different cores is about 0.02~K, i.e., even assuming a very low $T_{\rm dust}$ of $\sim$5~K, it is still about a factor of 250 lower. Such a big difference can not be explained purely by a beam filling effect, since it requires a small source size of $\lesssim$ 20~AU (a factor of 250 smaller than the spatial resolution, i.e., around 4.5\arcsec{}, or $\sim$ 4200~AU in Band 7). Thus it is more likely due to a small optical depth, i.e., $\tau \ll$ 1 in the observed bands, at least averaged on the core scale, although a small inner region that is optically thick is still possible in some cores. These results motivate future work on radiative transfer models of protostellar cores to predict these Band 6 to Band 7 flux ratios.


\subsection{Spectral lines}
\label{sec:lines}

\begin{figure*}[ht!]
\epsscale{1.1}\plotone{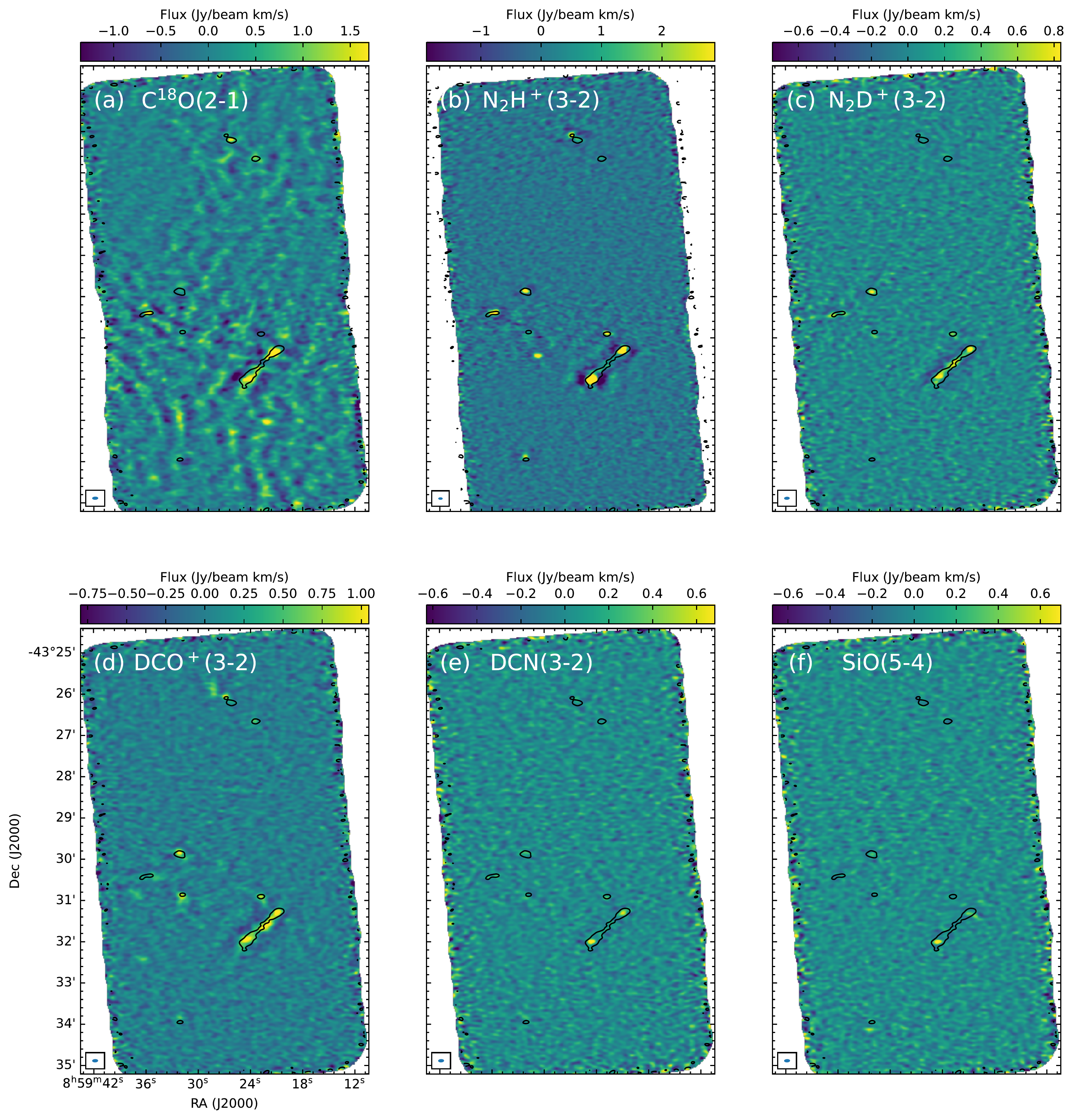}
\caption{Panels from (a) to (f) show the moment 0 maps of \ceighteeno(2-1), \ntwohp(3-2), \ntwodp{}(3-2),
\dcop{}(3-2), \dcn{}(3-2), and SiO(5-4). The $5\sigma$ 1.3mm continuum contour is overlaid in black for comparison.
}
\label{fig:mom0}
\end{figure*} 

\begin{figure*}[ht!]
\epsscale{1.1}\plotone{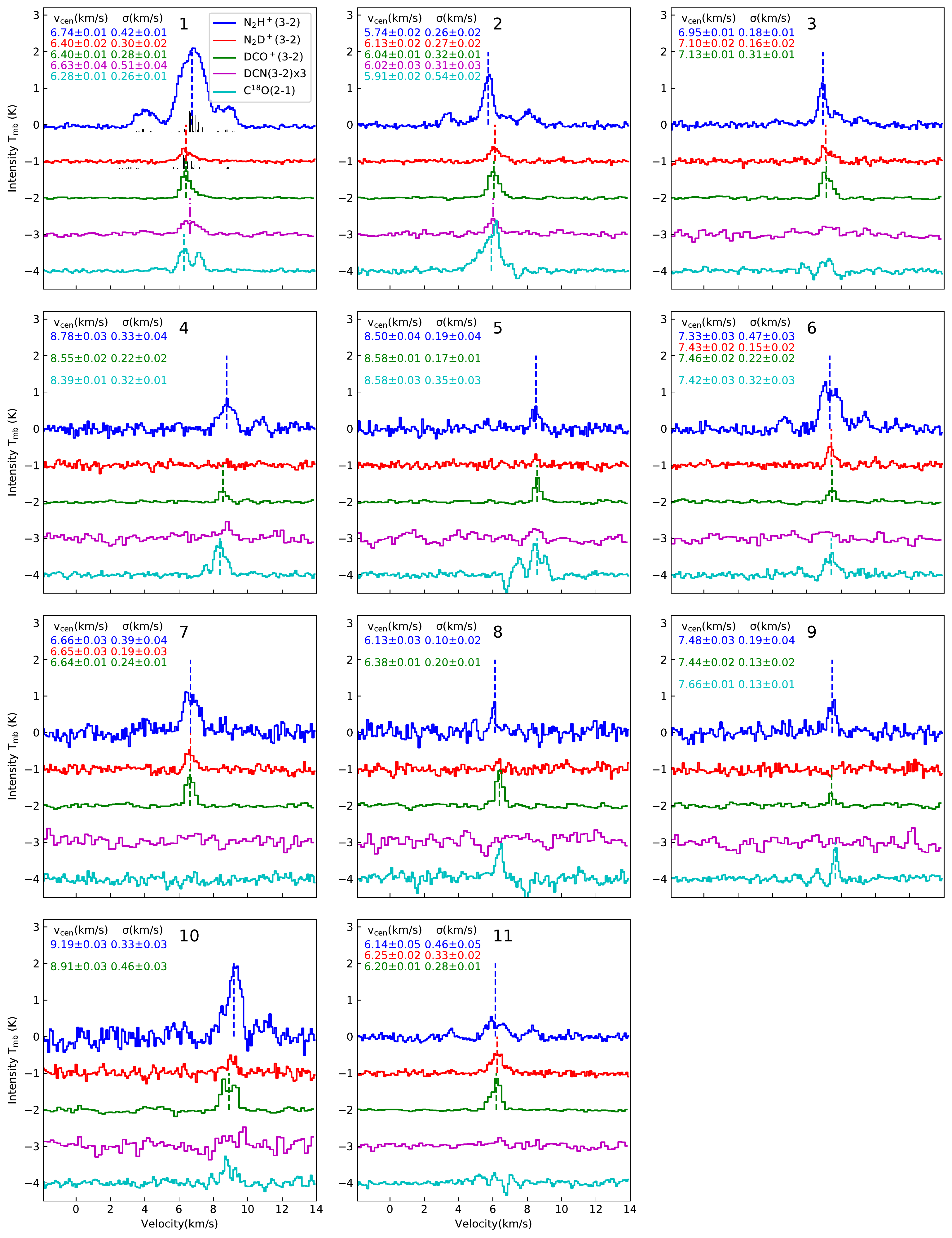}
\caption{Spectra (some with vertical offsets) of \ntwohp(3-2) (blue), \ntwodp(3-2) (red), \dcop{}(3-2) (green), DCN(3-2) (magenta) and \ceighteeno(2-1) (cyan) (see also legend in panel 1) of the 11 cores. In the first panel the relative intensities of hyperfine components of \ntwohp(3-2) and \ntwodp(3-2) are shown underneath the spectra. For spectra with a peak flux greater than $5\sigma$, we perform a Gaussian (or hyperfine profile weighted) fitting. The returned parameters (centroid velocity, velocity dispersion) for each line are displayed in the top left, in the same color as the corresponding line. The dashed vertical lines indicate the centroid velocities from line fitting. If there are multiple components for \ceighteeno(2-1), only the main component (the one closer to the other dense gas tracers; see text) is shown.}
\label{fig:spec}
\end{figure*} 

\autoref{fig:mom0} shows the moment 0 maps of \ceighteeno(2-1), \ntwohp(3-2), \ntwodp(3-2), \dcop(3-2), \dcn(3-2) and SiO(5-4). 
Other transitions described in \autoref{sec:observations} (\dcop(4-3), \dcn(4-3), $\rm{CH_3OH}(5_{1,4}-4_{2,2})$) do not have detection above 5$\sigma$ and hence are not included here. The maps of both \thirteenco(2-1) and \ceighteeno(2-1) appear strongly affected by missing large scale information due to the interferometric nature of the observations. It is likely that there exists significant CO line emission from nearby regions that are outside of the field of view, which hinders the performance of the cleaning process, and leads to strong sidelobes. In light of this we only include \ceighteeno(2-1) here for quantitative analysis, which is more optically thin and relatively less affected. \ntwohp{}(3-2) has strong detections and appears closely associated with the dust continuum. \dcop{}(3-2) is also associated with the dust continuum but slightly more extended. \ntwodp{}(3-2) and \dcn{}(3-2) have more limited detection compared with \ntwohp{}(3-2), and are only seen clearly towards a few cores. SiO(5-4) is only detected at the position of CR1c1, possibly tracing shocks related with accretion or outflows.

\begin{figure*}[ht!]
\epsscale{1.0}\plotone{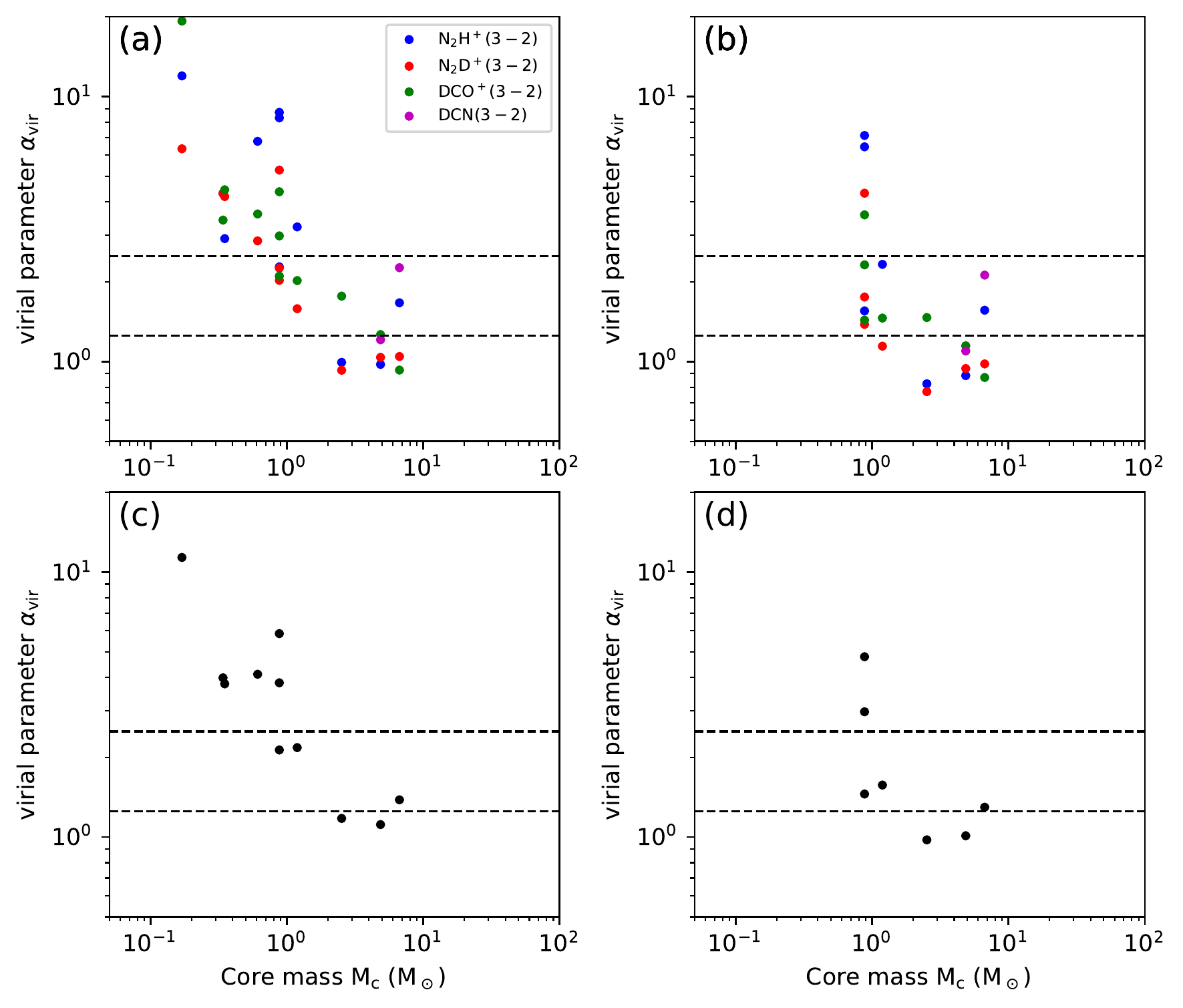}
\caption{
(a) Virial parameter, $\alpha_{\rm vir}$, versus core mass, $M_c$, with core radius measured
from the dendrogram defined area and velocity dispersion measured with
different dense gas tracers, as shown in the legend. The simple critical
value of $\alpha_{\rm vir,cr}=2a\rightarrow 2.5$ (see text) is shown by the upper
dashed line: cores below this line are gravitationally bound. The
lower dashed line shows the simple virial equilibrium case of
$\alpha=a\rightarrow5/4$.
(b) As (a), but with core radius estimated after allowing for
beam deconvolution. Small cores, i.e., with areas $<1.5A_{\rm beam}$
are excluded. (c) Same as (a), but we take the linear
average of the non-thermal line width measured via different tracers
to derive an average virial parameter. 
(d) Same as
(c), but using the deconvolved size.}
\label{fig:vir}
\end{figure*}

To investigate the kinematic and dynamical properties of cores, we extract the average spectra of each core, as shown in \autoref{fig:spec}. Among the four tracers, \ntwohp{} and \dcop{} have clear detections for almost all cores, while other lines are relatively weak and only detected for part of the core sample. The \ceighteeno{} profiles appear to be relatively complicated for some cores, like CR1c3 and CR1c5.
To measure the centroid velocity and velocity dispersion of each core
we perform a fitting on spectra with well defined profiles, i.e., those with a
peak greater than a certain threshold value. Here we adopt a
5$\sigma$ criterion for this threshold. Since the noise levels
of the average spectra vary for different cores (depending on the
pixel numbers in the core, etc.), we estimate the rms noise separately
for each core and each line using the signal-free channels. This
signal-to-noise criterion gives 6 cores for analysis with
\ceighteeno{}(2-1), 11 for \ntwohp{}(3-2), 6 for \ntwodp{}(3-2), 11 for
\dcop{}(3-2) and 2 for \dcn{}(3-2).

We characterize the \ceighteeno{}(2-1) spectra with Gaussian
fitting using the {\it curve\_fit} function in the {\it Scipy.optimize} python module, i.e., the brightness temperature at velocity $v$, $T_B(v)$, is given by
\begin{equation}
     T_B(v)= T_0  {\rm exp} \left[ - \frac{(v-v_{\rm cen})^2}{2\sigma^2} \right], 
\end{equation}
where $T_0 \simeq \tau_{0} T_{\rm ex}$ when the line is optically thin.
CR1c2 and CR1c6 can be well described with
a single Gaussian component. In general, we expect that
\ceighteeno{}(2-1) traces somewhat lower density envelope gas surrounding the
dense core and thus could be more affected by multiple components
along the line of sight. In CR1c1, CR1c4, CR1c5 and CR1c9, where the spectra have more complex
profiles and hence cannot be well approximated by a single Gaussian, we also allow 
for a second Gaussian component.
The component closest to the velocity determined from other dense gas tracers is assumed 
to be associated with the core. For the \dcop{}(3-2) and \dcn{}(3-2) lines we also perform the Gaussian fitting with the {\it curve\_fit} function. 

On the other hand, \ntwohp{} and \ntwodp{} lines have blended hyperfine components and cannot be approximated with a simple Gaussian. We adopt the frequencies and relative optical depths of \ntwohp{} and \ntwodp{} taken from \citet{Pagani09}.
We further assume the line emission is optically thin to limit the number of free parameters, i.e., 
\begin{equation}
     T_B(v)= T_0 \sum_i R_i  {\rm exp} \left[ - \frac{(v-v_i-v_{\rm cen})^2}{2\sigma^2} \right],
\end{equation} 
where $R_i$ and $v_i$ are the relative intensity and velocity for the ith hyperfine component, respectively.
For CR1c1 the signal to noise ratio is very high ($\sim$20) and three hyperfine groups are clearly 
detected, so we attempted to include the excitation temperatures ($T_{\rm ex}$), and opacities ($\tau_{\rm tot}$)
as free parameters to fit the profile, which is described in \autoref{sec:appA}.
The best-fit parameters of centroid velocity and velocity dispersion are displayed along with the spectral
lines in \autoref{fig:spec}. As can be seen, the centroid velocities range from 5.7 to 9.2 \kms{} and the velocity
dispersions $\sigma_{\rm obs}$ range from 0.15 to 0.5 \kms{} for all species, in which the nonthermal component can be derived via
\begin{equation}
            \sigma_{\rm nth}^2 = \sigma_{\rm obs}^2 - \sigma_{\rm th,obs}^2 = \sigma_{\rm obs}^2 - \frac{k T}{\mu_{\rm obs} m_{\rm H}}
\end{equation}
where $\mu_{\rm obs}$ is the mass of the particular tracer species. At a temperature of 15~K, the thermal dispersion $\sigma_{\rm th} = \sqrt{k T/\mu m_{\rm H}}$ is 0.23~\kms, with $\mu$ = 2.33 assuming $n_{\rm He}=0.1 n_{\rm H}$. Thus the Mach number is measured to range from 0.61 to 2.2, with a median of 1.4 for \ntwohp, 0.77 for \ntwodp{} and 1.0 for \dcop. 

\subsection{Virial state of cores}
\label{sec:vir}
To further examine the dynamical state of the dense cores, we calculate the virial parameter \citep{Bertoldi92}, defined as
\begin{equation}
 \alpha_{\rm vir} \equiv 5\sigma_c^2R_c/(GM_{c}) = 2aE_K/|E_G|,
\end{equation}
where $\sigma_c$ is the intrinsic 1D velocity dispersion of the 
core and $R_c$ is the core radius. The dimensionless parameter $a$ accounts for modifications that apply in the case of non-homogeneous and non-spherical density distributions and
we adopt a fiducial value of $a=5/4$ following \citet{Mckee03}, which corresponds to
a radial density profile of $\rho\propto r^{-1.5}$.
For a self-gravitating, unmagnetized core without rotation, a virial
parameter above a critical value $\alpha_{\rm vir,cr}= 2a$ indicates that 
the core is unbound and may expand, while one below $\alpha_{\rm vir,cr}$ 
suggests that the core is bound and may collapse. 

Following the procedures in \citet{Cheng20}, we calculate the virial parameters separately using each tracer, 
i.e., \ntwohp{}, \ntwodp{}, \dcop{} and \dcn. 
The intrinsic velocity dispersion $\sigma_c$ is derived from the observed dispersion $\sigma_{\rm obs}$ following:
\begin{equation}
\sigma_c =  \left(\sigma_{\rm nth}^2+\sigma_{\rm th}^2\right)^{1/2} \\
             =  \left(\sigma_{\rm obs}^2-\frac{k T}{\mu_{\rm obs} m_{\rm H}}+\frac{k T}{\mu m_{\rm H}}\right)^{1/2},
\end{equation}
 where $\mu$ = 2.33 is the mean molecular weight assuming $n_{\rm He}=0.1 n_{\rm H}$ and $\mu_{\rm obs}$ is the molecular weight of different observed species. For the core masses, we use the values from \autoref{table:core_property}, i.e., derived based on millimeter continuum emission. 
For core radius, we attempt two methods. The first is to use the effective radius calculated from the dendrogram-returned 
area in \autoref{sec:cont}.
For the second method, we adopt a deconvolved size defined as $R_c$ =  $\sqrt{(A - A_{\rm beam})/\pi}$
for cores with $A> 1.5 A_{\rm beam}$, where $A$ and $A_{\rm beam}$ are the core area and synthesized beam size, respectively. 
\autoref{fig:vir}(a) and (b) display the virial parameters measured with different tracers versus core mass for the two methods described above. In \autoref{fig:vir}(c) and (d), we combine the measurements from different tracers by taking the linear average of their nonthermal velocity dispersion in the virial parameter derivation.  We note that this procedure may potentially introduce some bias, since cores can vary in the number and type of chemical species that have detected line emission.




We see virial parameters ranging from 1 to 20 as measured by
individual dense gas tracers. There is a trend for more massive
cores to have smaller virial parameters. The scatter is
reduced for the deconvolved size method, suggesting some data points with 
virial parameter $>5$ in panel (a) could arise from an overestimation of the
core radius. There are no significant systematic differences between
different tracers. For example, with the deconvolved size method (panel (b)),
the median values are 1.56, 1.14 and 1.46 for \ntwohp, \ntwodp{} and \dcop. 
The virial parameters estimated by averaging all the 
available dense gas data for each core show a further reduction in the scatter. For the second method with deconvolved sizes that focus on the larger cores, we obtain a median value of 1.45, with 2 out of the 7 cores exceeding the critical value of $\alpha_{\rm vir,cr}$ = 2.5. For comparison, the virial parameters of the cores appear to be similar to those of the 76 cores in G286, which have a median value of 1.22 \citep{Cheng20}. Thus we see that most cores have a virial parameter that is consistent with a value expected in virial equilibrium. Note that the derivation of virial ratios relies on the assumption of temperature, which strongly affects the mass estimation. 
Here an uniform temperature of 15~K has been assumed. If we use a temperature of 20~K, then the median virial parameter rises to 2.58. 

Following similar discussions in \citet{Cheng20}, 
the absolute uncertainties in the derived virial parameters, including uncertainties 
in measured 1D line dispersion, mass and temperature, can be as high as a factor of 2.5.
Therefore, it is difficult to be more certain about whether the
dense cores are actually closer to a supervirial or subvirial state.
For example, CR1c11 has the highest virial parameter of 4.8, but if a lower temperature of 10 K is adopted, then $\alpha_{\rm vir}$ = 2.3, i.e., below the critical value of 2.5, so it is still likely to be gravitational bound. However, we note that the uncertainty factor includes systematic effects, some of which are 
not expected to vary that much from core to core, so the cores with smallest virial
parameters, like CR1c1, CR1c2 and CR1c3, are more likely to be gravitationally bound and collapsing.

\begin{figure}[ht!]
\epsscale{1.0}\plotone{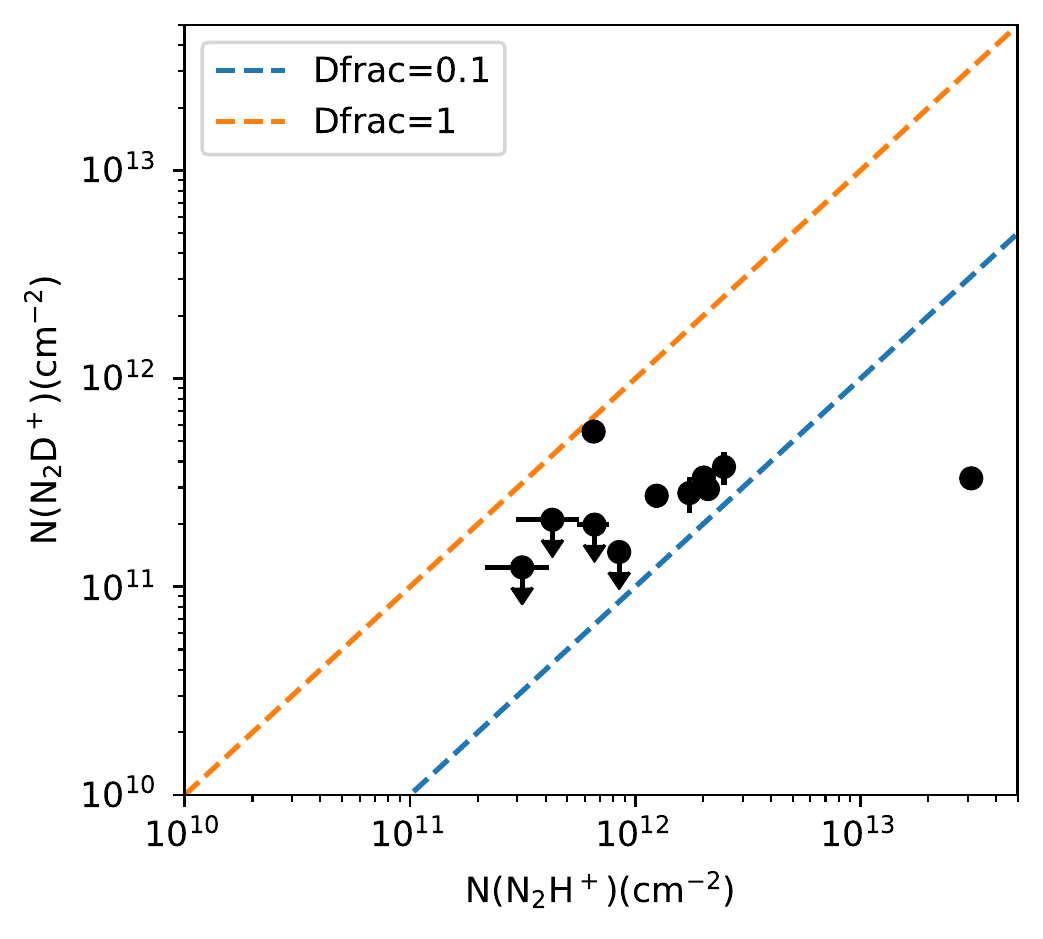}
\caption{Measured \ntwohp{} and \ntwodp{} column densities for the dense core sample. The two dashed lines are reference lines for \Dfrac = 0.1 and 1, respectively.}
\label{fig:deu}
\end{figure} 

\begin{deluxetable*}{ccccccccc}
\tabletypesize{\scriptsize}
\renewcommand{\arraystretch}{0.9}
\caption{Estimated column densities, deuteration fractions, CO depletion factors and infrared detections for the cores.}
\label{table:core_abun}
\tablehead{
\colhead{core} & \colhead{$N_{\rm H}$}& \colhead{$N({\rm C^{18}O})$}  & \colhead{$N({\rm N_2H^+})$} & \colhead{$N({\rm N_2D^+})$} & \colhead{\Dfrac} & \colhead{$f_D$} & \colhead{12~$\mu$m} & \colhead{70~$\mu$m} \\
\colhead{} &\colhead{($\rm 10^{22} cm^{-2}$)} & \colhead{($\rm 10^{14} cm^{-2}$)} & \colhead{($\rm 10^{11} cm^{-2}$)}& \colhead{($\rm 10^{11} cm^{-2}$)} & \colhead{}&\colhead{} &\colhead{} & \colhead{}
}
\startdata
1  &  9.75  &  4.79  &  310.57  &  3.32  &  0.011  &  62.4 & Y&Y\\ 
2  &  10.36 & 9.54  &  20.18  &  3.35  &  0.17  &  33.3 &Y &Y\\ 
3  &  9.48  &  1.00  &  12.46  &  2.74  &  0.22  &  290.7 &N &Y\\ 
4  &  6.87  &  4.79  &  8.49  &  $<$1.47  &  $<$0.17  &  43.9 &Y &Y\\ 
5  &  5.68  &  3.72  &  3.15  &  $<$1.24  &  $<$0.39  &  46.8 &N &N\\ 
6  &  4.19  &  3.39  &  21.05  &  2.95  &  0.14  &  37.8 &Y &Y\\ 
7  &  5.13 &$<$1.42  &  17.40  &  2.82  &  0.16  &  $>$110.7 &N &Y\\ 
8  &  4.16  &  2.87  &  4.29  &  $<$2.10  &  $<$0.49  &  44.5 &N &Y\\ 
9  &  4.02  &  1.24  &  6.60  &  $<$1.99  &  $<$0.30  &  99.2&Y &Y\\ 
10  &  3.72  &  2.04  &  24.80  &  3.77  &  0.15  &  55.8 &N &N\\ 
11  &  3.50  &  1.62  &  6.54  &  5.57  &   0.85  &  66.2 &N &N
\enddata
\end{deluxetable*}

\subsection{Deuteration and CO depletion}

\begin{figure*}[ht!]
\epsscale{1.0}\plotone{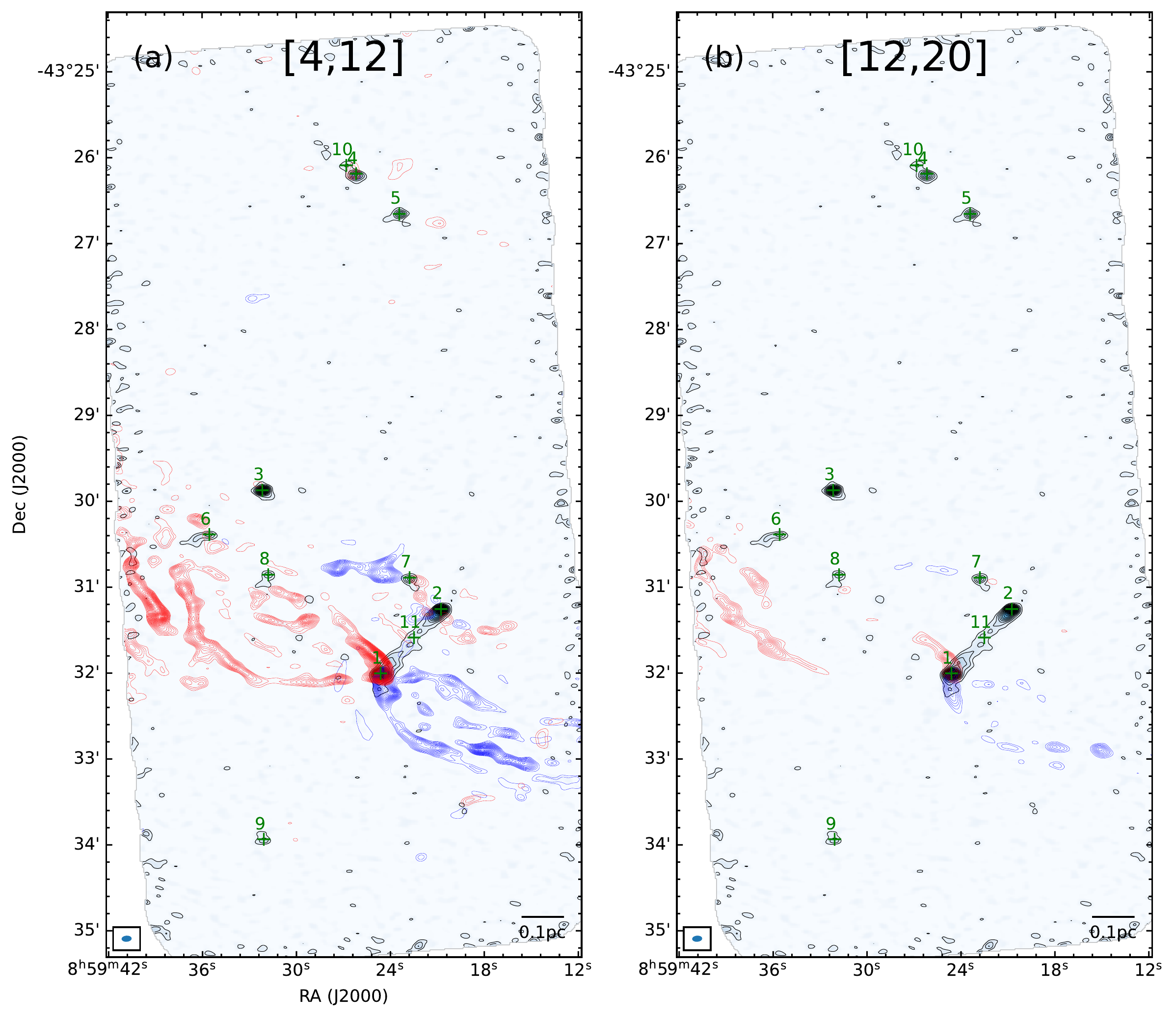}
\caption{(a) CO(2-1) emission integrated from relative velocities from -4 to -12~\kms for blueshifted and +4 to +12~ \kms{} for redshifted channels. The continuum is shown in grey scale and black contours for comparison.{\it(b)} CO(2-1) emission integrated from relative velocities from -12 to -20~\kms{} for blueshifted and +12 to +20~\kms for redshifted channels. The velocity ranges (relative to an averaged system velocity of 7 \kms ) are indicated on top of the panels.}
\label{fig:outflow}
\end{figure*} 

For optically thin lines, following \citet{Mangum15}, the column density is calculated from the line integrated intensity by
\begin{equation}
\begin{split}
    N_{\rm tot}^{\rm thin} = \left(\frac{3h}{8\pi^3S\mu^2_{\rm dm} R_i}\right) \left(\frac{Q_{\rm rot}}{g_J g_K g_I}\right) \frac{{\rm exp}(\frac{E_u}{kT_{\rm ex}})}{{\rm exp}(\frac{h\nu}{kT_{\rm ex}})-1}\\ \times
    \frac{1}{(J_\nu(T_{\rm ex})-J_\nu(T_{\rm bg}))}\int\frac{T_B dv}{f}
\end{split}
\end{equation}
where $J_\nu(T)\equiv\frac{h\nu/k}{{\rm exp}(h\nu/[kT])-1}$; $S$ is the transition line strength; $\mu_{\rm dm}$ is the molecular dipole moment, $R_i$ is the relative transition intensity (for hyperfine transitions), $Q_{\rm rot}$ is the rotational partition function, $T_B$ is the measured brightness temperature; 
$f$ is the filling factor, and $g_J$, $g_K$ and $g_I$ are the rotational degeneracy, K degeneracy and nuclear spin degeneracy, respectively. In our calculations we assume a fiducial excitation temperature of 10~K, i.e., moderately cooler than the fiducial dust temperature of 15~K. Such sub-thermal excitation conditions are motivated in part by the results of \citet{Kong16} who derived and/or considered excitation temperatures from about 4 to 7~K for $\rm N_2D^+$ and $\rm N_2H^+$ in massive cores in Infrared Dark Clouds (IRDCs), with these being significantly lower than gas temperature estimates of $\sim 10$ to $15\:$K from $\rm NH_3$ observations of the same regions \citep{Kong18}. Since our observations of $\rm N_2D^+$ and $\rm N_2H^+$ are mostly in protostellar core envelopes that are smaller-scale, denser and warmer than the IRDC regions studied by \citet{Kong16}, we consider $T_{\rm ex}=10\:$K to be the most appropriate fiducial choice. However, we will discuss, below, the effects of variation of this choice.




For the derivation of the column densities, we assumed a unity filling factor for all sources. The column density of different species are summarized in \autoref{table:core_abun}. 
For the \ntwohp{} line emission of CR1c1, since the opacity can be determined from the spectral line fitting, the column density is corrected by
\begin{equation}
    N_{\rm tot} = N_{\rm tot}^{\rm thin}\frac{\tau}{1-{\rm exp}(-\tau)}.
\end{equation}
Furthermore, with the derived column densities the deuteration ratio for each core is estimated as $D_{\rm frac}=N({\rm N_2D^+})/N({\rm N_2H^+})$. The results are listed in \autoref{table:core_abun}. \autoref{fig:deu} shows the \ntwohp{} and \ntwodp{} column density measurements of the dense cores. The \ntwohp{} column densities are in the range of $\rm 3 \times 10^{11}$ - $\rm 3 \times 10^{13} cm^{-2}$, while the \ntwodp{} column densities are in the range $\rm 10^{11}$ - $\rm 6 \times 10^{11} cm^{-2}$. The values of \Dfrac{} are between 0.011 and 0.85, with a median value of 0.16.
This is similar to the value found by \citet{Crapsi05} in their sample of low-mass starless cores.

The uncertainties in the column density estimation mainly result from the assumption of the excitation temperature $T_{\rm ex}$. If temperatures of 7~K or 15~K were adopted, then $N(\rm N_2H^+)$ would vary by factors of 2.3 and 0.59, respectively, and $N(\rm N_2D^+)$ would vary by factors of 1.9 and 0.69, respectively. Nevertheless, assuming the species have the same excitation temperature, the deuteration ratio $D_{\rm frac}$ is relatively robust and differs only by factors of 0.83 to 1.17 from the low to the high temperature limits of this range.
The uncertainties in flux measurement are typically $<$ 10\% for \ntwohp, with only a few exceptions for the cores with weaker \ntwohp{} emission, i.e., CR1c5, CR1c8 and CR1c9, that have $\lesssim$ 30\% uncertainties. For \ntwodp{} the uncertainties in flux measurement are all $<$ 20\%. Additionally, there are flux calibration uncertainties of about 10\% for Bands 6 and 7, respectively. 

The CO depletion factor, $f_D$, is defined as the ratio between the ``expected'' abundance of CO and the ``observed'' value:
\begin{equation}
    f_D = \frac{X_{\rm C^{18}O}^{\rm exp}}{X_{\rm C^{18}O}^{\rm obs}}
\end{equation}
In the abundance calculation we derive the column density of hydrogen nuclei, $N_{\rm H}$ from the mass surface density $\Sigma_c$ listed in \autoref{table:core_property} by $N_{\rm H}$ = $\Sigma_c/\mu_{\rm H} m_{\rm H}$, where $\mu_{\rm H} m_{\rm H} = 1.4 m_{\rm H}$ is the mean mass per H nucleus. To compute $X_{\rm C^{18}O}^{\rm exp}$ we adopt the abundance ratios of $n_{\rm 16O} / n_{\rm 18O}$ = 327 from \citet{Wilson94} and $n_{\rm 12CO}/n_{\rm H_2}$ = 2$\times 10 ^{-4}$ from \citet{Lacy94}. Thus, our assumed abundance ratio of \ceighteeno{} to \htwo{} is 6.12 $\times10^{-7}$. The results are listed in \autoref{table:core_abun}. All the cores have $f_D$ measured to be $\gtrsim$ 40. 
 Note that the imperfect cleaning due to incomplete uv sampling may have affected the \ceighteeno{} flux measurement, and CO depletion factor accordingly. As mentioned in \autoref{sec:lines} the moment 0 map of \ceighteeno(2-1) in \autoref{fig:mom0} does have some artificial ringing features and some cores are not clearly associated with enhanced \ceighteeno{} emission. It is difficult to quantify the uncertainties introduced from the cleaning process, but it may have affected the CO depletion factor by factors of a few for specific cores.



\begin{figure*}[ht!]
\epsscale{1.2}\plotone{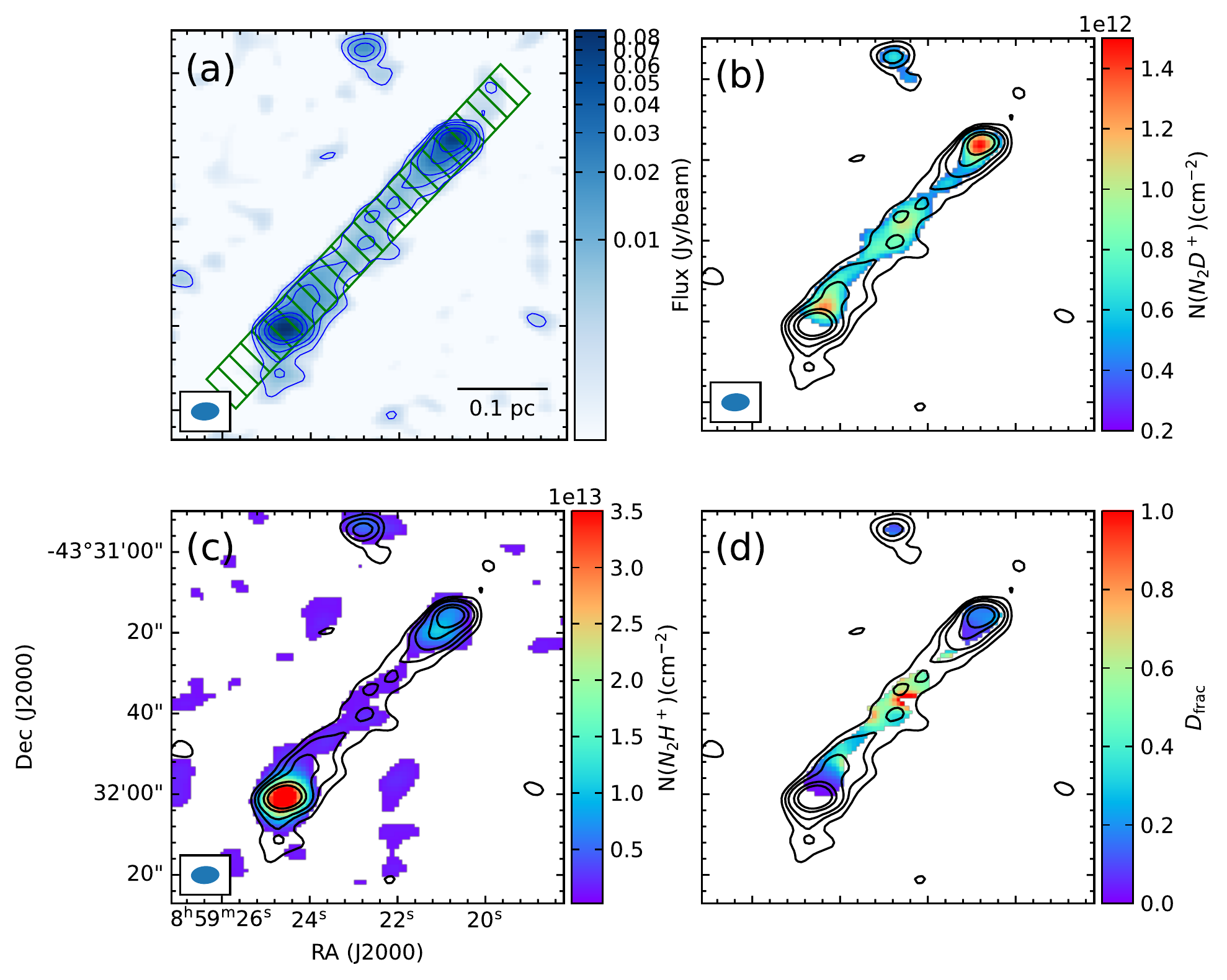}
\caption{{\it (a)} 1.3~mm continuum map of the bridge feature between the two most luminous cores shown in blue color scale and contours. The contour levels are $\sigma$ \  $\times$ (5, 10, 15, 30, 50), with 1$\sigma$ = 1.3 \mjypbm. As shown in green rectangles we have divided this region into 20 blocks to extract properties along the bridge feature. See text for more details.  {\it (b)} \ntwodp{} column density map. The 1.3~mm continuum is overlaid for comparison. {\it (c)} \ntwohp{} column density map. The 1.3~mm continuum is overlaid for comparison. {\it (d)} \Dfrac{} map. The 1.3~mm continuum is overlaid for comparison.
}
\label{fig:bridge}
\end{figure*} 

\begin{figure*}[ht!]
\epsscale{1.1}\plotone{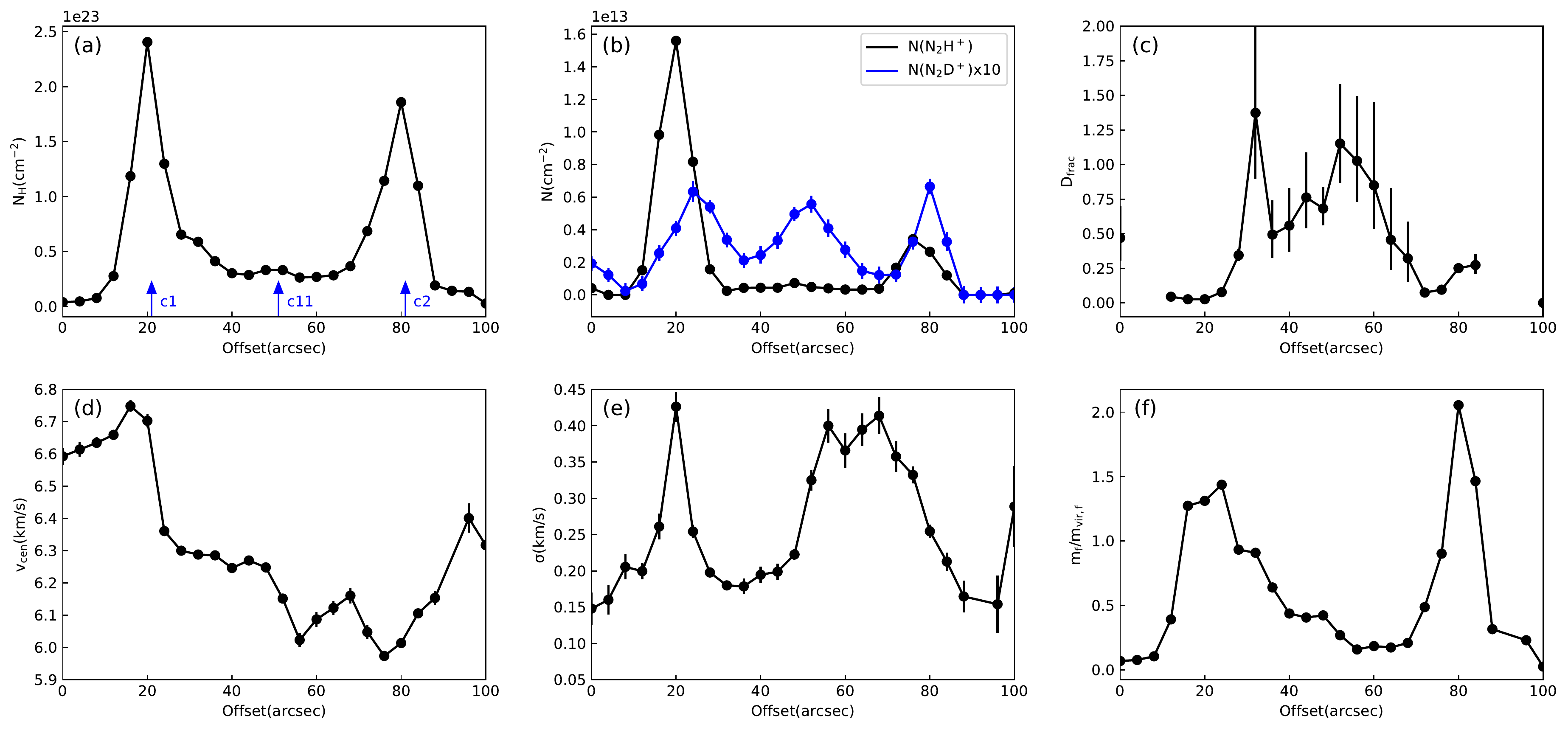}
\caption{Measured properties along the filamentary bridge feature. {\it (a)} Column density of H nuclei, $N_{\rm H}$. The positions of three cores (CR1c1, CR1c2, CR1c11) are indicated by blue arrows. 
{\it (b)} Column density of \ntwohp{} and \ntwodp{} are shown by black and blue points/lines, respectively.  The column density of \ntwodp{} is enlarged by a scaling factor of 10 for ease of viewing.
{\it (c)} Deuteration fraction of \ntwohp{}, \Dfrac{}.
{\it (d)} Centroid velocity measured with the averaged \dcop{} spectrum of each block.
{\it (e)} Velocity dispersion measured with the averaged \dcop{} spectrum of each block.
{\it (f)} 
Ratio of mass per unit length to virial mass per unit length, $m_f/m_{\rm f,vir}$. The mass per unit length, $m_f$, is calculated from the 1.3~mm continuum, while the virial mass per unit length, $m_{\rm f,vir}$, is derived with the velocity dispersion measured from the \dcop{} spectra. See text for more details.
}
\label{fig:bridge_stat}
\end{figure*}

\subsection{CO Outflows}

We examined the CO(2-1) data toward this region to see if protostellar outflows are detectable. \autoref{fig:outflow}(a) illustrates the low velocity CO(2-1) emission integrated over relative velocities ranging from 4 to 12~\kms{} (compared to $v_{\rm sys}\approx 
7$~\kms) for blueshifted and redshifted emission, and \autoref{fig:outflow} (b) illustrates the high velocity CO(2-1) emission integrated over relative velocities ranging from 12 to 20~\kms{}.
There is a clear bipolar outflow associated with CR1c1, which has an orientation roughly perpendicular with the filamentary bridging feature seen in the continuum. The outflow has a biconical shape with an half opening angle of $\sim$30$^{\circ}$. 
In the vicinity of CR1c2 there appears to be some blueshifted and redshifted CO emission, possibly resulting from a weak outflow, which is also perpendicular to the bridging filament.
CR1c7 appears to host a relatively collimated outflow in East-West orientation. The blueshifted lobes has a knotty appearance with a bending feature extending to Northeast direction.  There is also a tentative detection of CO outflow from CR1c4 at relatively low velocities in the redshifted lobe, suggesting CR1c4 may also host a protostar.

We also examined the CO channel maps centered on CR1c3, CR1c5, CR1c6, CR1c8, CR1c9, CR1c10 and CR1c11 and did not find evidence for outflows. The strong CO emission from the molecular cloud and the spatial filtering, however, make these nondetections questionable, and observations with higher signal-to-noise are required to properly establish the presence or lack of CO outflows from these sources.

\subsection{The bridging filament connecting cores CR1c1 and CR1c2}

In the continuum map there is an interesting linear filament in which CR1c1, CR1c2 and CR1c11 are located. CR1c1 and CR1c2 are located at the ends of this filament and connected by extended emission seen in 1.3~mm continuum. CR1c11 lies in between CR1c1 and CR1c2 and is further identified from the moment 0 maps of \ntwodp, \dcop{} and 1.05~mm continuum. These three cores exhibit
signatures of different evolutionary stages: both CR1c1 and CR1c2 are associated with outflows and have relatively larger values of \fratio{} (1.83, 1.59, respectively), indicating that they already host a protostar that is actively accreting and heating up the surroundings. CR1c11 shows no sign of star formation activity and has a low value of \fratio{} of 1.28. As discussed in \autoref{sec:vir}, CR1c11 could be gravitationally bound if a lower temperature of $\lesssim$ 10~K is assumed. If true, then CR1c11 may be a prestellar core. The chemical properties including \Dfrac{} are also consistent with these differences in evolutionary stage. 

We further divide the bridging filament into 20 strips to derive properties along its length, as shown in \autoref{fig:bridge}. Each strip has a size of 7\arcsec $\times$ 3.5\arcsec. The column densities of \ntwohp{}, \ntwodp{} are calculated following the procedures in \autoref{sec:lines} and also shown in \autoref{fig:bridge}.
 Note that in addition to the uncertainties discussed in \autoref{sec:lines}, spatially filtering of interferometer observations may also lead to an underestimation of flux measurements along the bridge. For example, if there is a more diffuse cocoon component surrounding the bridge we are probably not able to detect it with the current observations. The hydrogen column density $N_{\rm H}$ is calculated from the continuum emission assuming a uniform $T_{\rm dust}$ of 15~K as in \autoref{sec:cont}. 
We plot the derived column densities, as well as \Dfrac{} in \autoref{fig:bridge_stat}. 
The evolutionary differences are
better illustrated in the \Dfrac{} profile, which exhibits a plateau around \Dfrac{} $\approx$ 0.8 from 30\arcsec{} to 60\arcsec{}, i.e., covering the bridging region between CR1c1 and CR1c2. It can also be seen that CR1c1, CR1c2 and CR1c11 have similar $N$(\ntwodp), but there is a lack of \ntwohp{} for CR1c11, thus leading to a high \Dfrac. Therefore, CR1c11 is expected to be in an early stage before the onset of star formation. 


To investigate the kinematic properties, we check the line spectra along the bridging filament. \dcop{}(3-2) is the best tracer for this purpose, since it is clearly detected throughout the bridge and has a better signal to noise ratio compared to \ntwodp(3-2).
We fit the \dcop{} spectra with the same routine as used in \autoref{sec:lines}. \autoref{fig:bridge_stat} illustrates the variation of centroid velocity and velocity dispersion along the bridge. The \dcop{} velocity dispersion ranges from 0.15 to 0.45 \kms. 
With gas temperatures of 10-20 K, the 
thermal line broadening is 
0.05-0.07 \kms{} for \dcop, so the observed line width is dominated by the nonthermal component.
The thermal sound speed of molecular gas is 0.23~\kms{} at 15~K,
and so the Mach number ranges from 0.6 to 2. The filament appears mildly subsonic at the relative quiescent part, i.e., at offsets from 30\arcsec{} to 50\arcsec. Note that there could be multiple velocity components along the filament that are unresolved in the current observation. There is a clear peak in line dispersion at the position of CR1c1, possibly resulting from an increase in temperature or enhanced nonthermal motions, such as infall and/or outflow due to star formation activity. The case of CR1c2 and CR1c11 is less clear. We see an increase in $\sigma_{\rm DCO^+}$ from 50\arcsec{} to 80\arcsec{} in offset, which is roughly in between CR1c2 and CR1c11. 

\begin{figure}[ht!]
\epsscale{1.0}\plotone{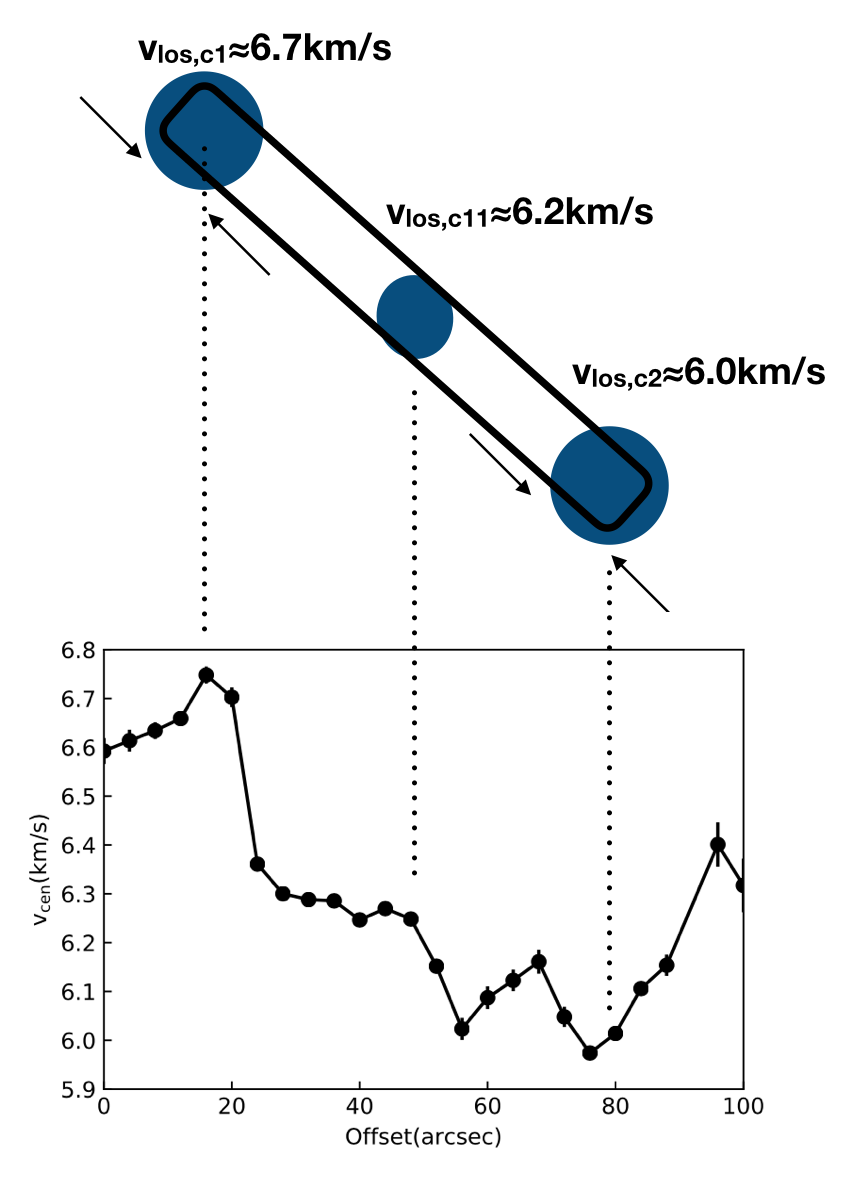}
\caption{Schematic diagram of a possible scenario to explain the centroid velocity profile in \autoref{fig:bridge_stat}. The observation is made from the bottom of this plot. The global velocity gradient between CR1c1 and CR1c2 may result from mechanisms like filament rotation, while the maxima or minima on the velocity profile are caused by local infall motions around CR1c1 and CR1c2.}
\label{fig:sketch_fila}
\end{figure}

For $v_{\rm cen}$ there is a decreasing trend from 6.7 \kms{} at 20\arcsec, to 6.0 \kms{} at around 80\arcsec, indicating a global velocity gradient of about 2.6 $\rm km s^{-1} pc^{-1}$. 
Velocity gradients along filaments have been observed in both nearby low-mass star-forming clouds \citep[e.g.,][]{Hacar11} and massive clouds \citep[e.g.][]{Henshaw13,Peretto14}, and often interpreted as flows along filaments, feeding gas into dense cores.
However, the global velocity gradient in the filaments may also be attributed to the motion of the filaments themselves (e.g., rotation or oscillation along the line of sight) rather than accretion flows.
Interestingly, the positions of CR1c1 and CR1c2 seem to coincide well with local maxima or minima on the $v_{\rm cen}$ profile, possibly suggesting gas infall is taking place in the vicinity of the cores. \autoref{fig:sketch_fila} illustrates a possible scenario to explain the observed $v_{\rm cen}$ variations, in which 
the local bending feature of the velocity profile is caused by infall and/or rotational motion around CR1c1 and CR1c2, while the global velocity gradient between CR1c1 and CR1c2 may arise from other mechanisms like rotation.
Summarizing the results, it is likely that the bridging feature is a remnant of a larger filament. CR1c1 and CR1c2 have been accumulating gas material from this filament and have formed protostars, while CR1c11 has condensed from the gas reservoir more recently and is still in a very early, starless evolutionary stage. 


To investigate the dynamic state of the filament we perform a filamentary virial analysis following \citet[][]{Fiege00}. As shown by \citet{Fiege00}, a pressure-confined, non-rotating,
self-gravitating, filamentary (i.e., length $\gg$ width) magnetized
cloud that is in virial equilibrium satisfies

\begin{equation}
\frac{P_e}{P_f} = 1 - \frac{m_f}{m_{\rm vir,f}}\left(1-\frac{M_f}{|W_f|} \right)
\end{equation}
where $P_f$ is the mean total pressure in the filament, $P_e$ is the
external pressure at its surface, $m_f$ is its mass per unit length,
$m_{\rm vir,f} = 2\sigma_f^2/G$ is its virial mass per unit length,
and $M_f$ and $W_f$ are the gravitational energy and magnetic energy
per unit length, respectively. Here, because of the observational
difficulties of measuring the surface pressure and magnetic fields, we
ignore the surface term and magnetic energy term, i.e., only
considering the balance between gravity and internal pressure support. 

The 100\arcsec\ length of the filament corresponds to 0.45~pc at an
assumed distance of 0.93~kpc. Without direct observational constraints, we further assume the
filament axis is inclined by an angle $i = 60^\circ$ to the line of
sight (90$^\circ$ would be in the plane of the sky). If an inclination
angle of 90 or 30$^\circ$ were to be adopted, then the length
estimates would differ by factors of 1.15 and 0.577,
respectively. Thus the actual length of the filament is assumed to
be 0.52~pc. 
In \autoref{fig:bridge_stat} we plot the ratio $m_{\rm f}/m_{\rm f,vir}$. The masses are calculated from the 1.3~mm continuum flux, assuming a temperature of 15~K and other dust properties as in \autoref{sec:cont}. 
$m_{\rm f,vir}$ is calculated using the velocity dispersion measured from \dcop{}.
The values of $m_f/m_{\rm vir,f}$ along the filament range from 0.2 to 2.0. $m_f/m_{\rm vir,f}$ clearly peaks at the positions of CR1c1 and CR1c2, with peak values of 1.4 and 2.0, respectively, and it is relatively small ($\sim$ 0.2-0.6) in regions between the two cores, suggesting that the filament may only be gravitationally bound around the positions of CR1c1 and CR1c2. 
However, since the ALMA 7m-array observations only probe scales up to $\sim$ 19\arcsec, they may be missing some flux from the filament leading to an underestimation of the masses. Furthermore, if a temperature of 10~K instead of 15~K is adopted, which is probably more realistic for the less evolved region between CR1c1 and CR1c2, the estimated mass will be larger by a factor of 1.85, thus bringing the $m_f/m_{\rm vir,f}$ ratio to $\sim$ 0.4-1.2.  
Also given other systematic uncertainties in measuring lengths of the structure, it is still likely that the majority of the filament is in approximate virial equilibrium, even without accounting for surface pressure and magnetic support terms.

\section{Discussion}
\label{sec:discussion}
\subsection{The dense gas fraction: a deficit in compact substructures}

An obvious feature in the continuum map of Vela C CR1 clump is an overall deficit of compact substructures at a few 0.01 pc scales. We have identified 11 cores from the 1.3 mm continuum, which add up to a total mass of only 19.4 $M_\odot$. Alternatively, if we sum up the fluxes above 4$\sigma$ in the 1.3 mm map and convert to masses following the same assumptions as in \autoref{sec:cont}, 
it yields 20.7 $M_\odot$, suggesting the bulk of the ALMA 1.3 mm emission is included in our identified dense cores. 
For comparison, the total clump mass in the field of view estimated from the {\it Herschel} column density map is about 2300 $M_\odot$, leading to a dense gas fraction, $f_{\rm dg}$, of only 0.84\% (or 0.90\%, using the total integrated flux). 
Therefore, only a very small fraction of gas mass is currently contained in 
compact prestellar and protostellar cores. 
The estimation of dense core masses depends on dust opacity, gas-to-dust mass ratio, temperatures and dust emission fluxes, as well as the distance to the region. The major uncertainty of mass estimation arises from the assumption of temperature. For example, if we assume a higher temperature of $T = 20$~K, the total mass will be a factor of 0.677 smaller, leading to a dense gas fraction of 0.57\% or 0.61\%. 
Note that for estimating $f_{\rm dg}$, some of these uncertainties cancel out, i.e., those due to distance and gas-to-dust mass ratio, so we expect the dense gas fraction in VelaC CR1 clump is $\lesssim 2\%$.


We compare the CR1 clump with another well studied region, G286.21+0.17 (G286), which is a protocluster at a distance of 2.5 kpc \citep{Cheng18}. G286 has a total {\it Herschel}-estimated mass of around 2900 $M_\odot$ in a 2.6'$\times$1.7' elliptical aperture \citep{Cheng20}, leading to an average column density, $N_{\rm H}$, of $\rm \sim 4 \times 10^{22} cm^{-2}$, 
similar to the Vela C CR1 clump ($\rm \sim 5 \times 10^{22} cm^{-2}$ ). For the compact gas mass we adopt two methods. For method 1 we simply sum up the masses of cores listed in \citet{Cheng20}, which follows the same assumptions as in \autoref{sec:cont}. 
For method 2 we integrate the fluxes for pixels above 4 $\sigma$ using the 1.3 mm continuum image made with only the 12m-array, and then convert to masses following the same assumptions. The 12m-array data of G286 have a maximum recoverable scale of 11\arcsec, corresponding to 0.13 pc at the distance of 2.5 kpc, which is close to the 7m-array observation of Vela C (sensitive to structures up to 29\arcsec~, $\sim$ 0.13 pc, in Band 6). Methods 1 and 2 yield $f_{\rm dg}$ of 7.3\%, and 14.3\% in G286, respectively, so both estimations are an order of magnitude higher compared with the Vela C CR1 clump. 

One possible explanation for these differences is that the formation of dense substructures in the Vela C CR1 clump has been suppressed by its strong magnetic field. Alternatively, the CR1 clump could simply be in a very early evolutionary stage of collapse, but with core formation not particularly influenced by the $B$-field. Follow-up observations to constrain the dynamical and chemical history of Vela C CR1, e.g., to measure infall speeds and chemical ages, can help distinguish these possibilities.

There have been a number of other studies of dense gas fractions in the literature. Direct comparison with our results is generally more difficult given the variety of methods used to estimate masses for both the large scale cloud and the dense (or compact) component.
For example, \citet{Battersby20} studied the dense gas fractions of the central molecular zone (CMZ) and compared to similar studies of clouds in the Galactic disk, finding that $f_{\rm dg}\sim 0.1\%$ to 2\% in most CMZ clouds (even though these clouds have relatively high column densities), while typical star-forming Galactic clouds have $f_{\rm dg}\sim 2\%$ to 20\%. The measured $f_{\rm dg}$ of Vela C CR1 clump appears similar to the CMZ clouds, and lower than typical Galactic disk clouds. But note that the maximum recoverable scale of our observation (29\arcsec, ~0.13pc) is smaller than the scales probed with SMA observations in \citet{Battersby20} for most sources.

\begin{figure}[ht!]
\epsscale{1.0}\plotone{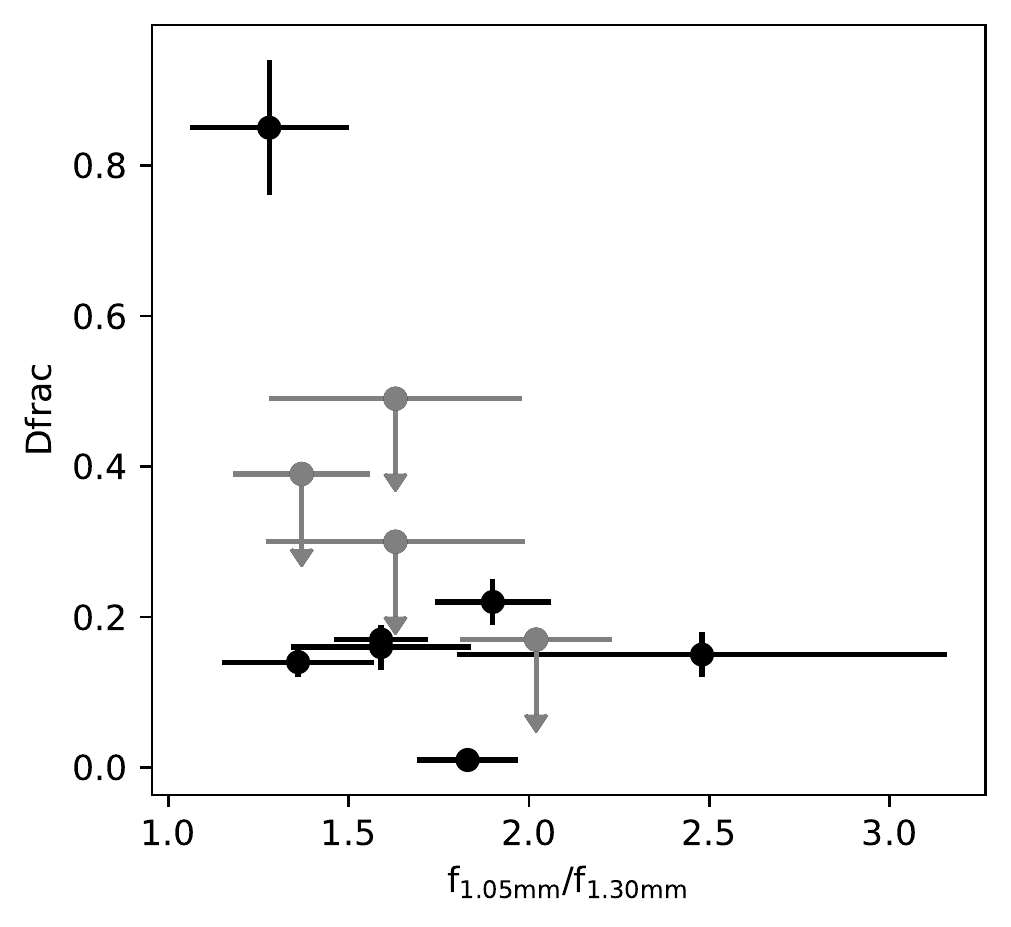}
\caption{Measured \Dfrac{} v.s. \fratio{} for the dense core sample. The data points that are upper limits are shown in grey.}
\label{fig:stat}
\end{figure} 

\subsection{Implication of the deuteration analysis: tests of astrochemical models}

The study of deuterated molecules is an important probe of the physical conditions in star-forming regions. Prior to the formation of a star, the cold ($T<$20~K) and dense ($n_{\rm H}>10^5\:{\rm cm}^{-3}$) conditions within star-forming molecular cloud cores drive a cold-gas chemistry that has been well studied in recent years. Many molecular species, including CO and its isotopologues, become depleted in the gas phase by freezing out onto dust grains. Unlike CO, N-bearing species, in particular \ammonia{} and \ntwohp, better trace dense and cold gas \citep[e.g.,][]{Caselli99,Bergin02}. This is due to the fact that CO, largely frozen out, is unable to effectively destroy their molecular ion precursors. These physical/chemical properties are commonly observed in prestellar cores, where the deuteration fraction (i.e., \Dfrac) of non-depleted molecules, defined as the column density ratio of one species containing deuterium to its counterpart containing hydrogen, is orders of magnitude larger than the average interstellar [D/H] abundance ratio, which is $\sim 10^{-5}$ \citep{Oliveira03}. Therefore, deuterated species, like \ntwodp{} are better suited to probe the physical conditions of the earliest stages of star formation. The \Dfrac(\ntwohp) ratio has been found to be a good evolutionary indicator in both low- and high-mass star formation \citep{Friesen10,Fontani11}. In addition, \ntwodp{} is probably the best tracer of prestellar cores, e.g., compared to \Dfrac{} of HNC and \ammonia \citep{Fontani15}.

The \Dfrac(\ntwohp) in the Vela C CR1 clump is found to be in the range of 0.011-0.85. 
Our observed values are consistent with measurements made in other low-mass star-forming regions \citep[e.g.,][]{Caselli02,Crapsi05,Daniel07,Emprechtinger09,Friesen13}. For 4 out of 11 cores, no significant \ntwodp{} is detected and only an upper limit of the \Dfrac{} is given. These cores also have relatively low \ntwohp{} column densities and the upper limit on \Dfrac{} ($\lesssim$0.5) is a rather loose constraint. The extreme value of 0.85, measured towards CR1c11, is among the highest levels of \ntwohp{} deuteration reported so far \citep[e.g.,][]{Miettinen12}, indicating the prestellar nature of CR1c11 in a very dense and cold condition. 
A caveat is that CR1c11 is defined based on the \ntwodp{} moment 0 map, which thus biases towards a higher \Dfrac{} estimation. 
We note that \ntwohp{} could have a greater degree of missing flux compared with \ntwodp{}, given the properties of the observations in Band 6 and Band 7, however, we do not expect significant flux losses on the scales of the observed cores.
In \autoref{fig:stat} we plot the \Dfrac{} ratio against other core properties to look for potential correlations. As discussed in \autoref{sec:cont}, the ratio \fratio{} can be interpreted as a temperature indicator, with higher \fratio{} suggesting higher temperature. There appears to be a weak anti-correlation between \fratio{} and \Dfrac{}, which is consistent with our expectation, since CO will be released from dust grains at higher temperatures as cores evolve, thus leading to a lower deuteration level.

Given the current available information on core properties it is difficult to assign a precise evolutionary stage for each one, but we do see groups of cores in different evolutionary stages. CR1c1 is probably the most evolved source in CR1. This core has the lowest \Dfrac{} and drives a powerful, wide-angle CO outflow. CR1c2, CR1c4 and CR1c7 also have associated outflow detections, indicating their protostellar nature. CR1c11 has the highest \Dfrac{} and is likely a prestellar core that is on the verge of collapsing, although more sensitive observations, especially better temperature measurements, are required to confirm its nature as a gravitationally bound core. A measurement of deuteration on the larger clump scale using single dish observations would be important for understanding the initial astrochemical conditions of prestellar core formation.

The auxiliary infrared data provide extra constraints on the evolutionary stages. Here we focus on two infrared wavelengths, i.e., 12~$\mu$m and 70~$\mu$m. A more complete investigation of other infrared wavelengths in presented in \autoref{sec:appB}. The 12~$\mu$m emission usually suggests thermal dust emission heated by protostars and 70~$\mu$m data could reveal deeply embedded protostars that are undetected at shorter wavelengths. The detection status is summarized in \autoref{table:core_abun}. As can be seen, CR1c5 and CR1c11 are not detected at 70~$\mu$m, suggesting that they are in a very early evolutionary stage, likely prestellar. The detection of CR1c10 is confused by another adjacent bright source (see \autoref{sec:appB}). All other cores should have formed a protostar. While CR1c3, CR1c7 and CR1c8 are detected at 70~$\mu$m, they are still very faint and not seen at 12~$\mu$ and hence they should be in a relatively earlier stage compared with other cores (i.e., CR1c1, CR1c2, CR1c4, CR1c6 and CR1c9), which are bright in both 12~$\mu$m and 70~$\mu$m and hence more evolved.


\section{Summary}
\label{sec:conclusion}

 The Vela C cloud is one of the few regions with magnetic field mapped through both sub-mm emission polarimetry and near-infrared stellar absorption polarimetry, and hence an ideal laboratory to study how the magnetic field strength affects star formation process. To investigate how star formation proceeds in a strong magnetic field environment, we have observed the Center Ridge 1 (CR1) clump in the Vela C with ALMA in Band~6 and Band~7. This clump is a high column density region that shows the lowest level of dust continuum polarization angle dispersion in the BLASTPol survey \citep{Fissel16}, indicating the presence of a strong magnetic field. We identified 11 dense cores via their mm continuum emission, with masses spanning from 0.17 to 6.7 {$M_\odot$}. Interestingly, CR1 exhibits a relatively low compact dense gas fraction compared with other typical clouds with similar column densities, which may be a result of the strong magnetic field in this region and/or that it is in a very early evolutionary stage of collapse.

 The \ntwohp(3-2) and \ntwodp(3-2) lines in this observation also allow for a precise measurement of the deuteration ratio. In our sample values of \Dfrac{} span from 0.011 to 0.85 for the dense core sample, with the latter being one of the highest values yet detected. A trend of decreasing \Dfrac{} from the final prestellar to protostellar phases is inferred by comparison to other indicators, such as presence of outflows and infrared sources. In addition we also report the detection of an bridging feature connecting the two most massive cores (CR1c1, CR1c2) in the region in both continuum and spectral lines. This linear filament is approximately parallel to the large scale plane of sky magnetic field orientation, and roughly orthogonal to the axes of CO bipolar outflows associated with CR1c1 and CR1c2. The kinematics of this filament likely imply that infall is occurring onto the cores.

 The presented study uses analysis methods for core identification and characterization from the ALMA Band 6 data that are the same as employed in studies of other star-forming regions, e.g., G286 by \citet{Cheng18,Cheng20} and IRDCs by \citet{Liu18,Liu20}. Future work will aim to extend such studies to other star-forming environments and thus allow a systematic investigation of many aspects of the star formation process and their dependence on galactic environment.

\appendix
\counterwithin{figure}{section}
\counterwithin{table}{section}
\section{\ntwohp(3-2) Spectral Fitting of of CR1c1}
\label{sec:appA}

\begin{figure}[ht!]
\epsscale{1.0}\plotone{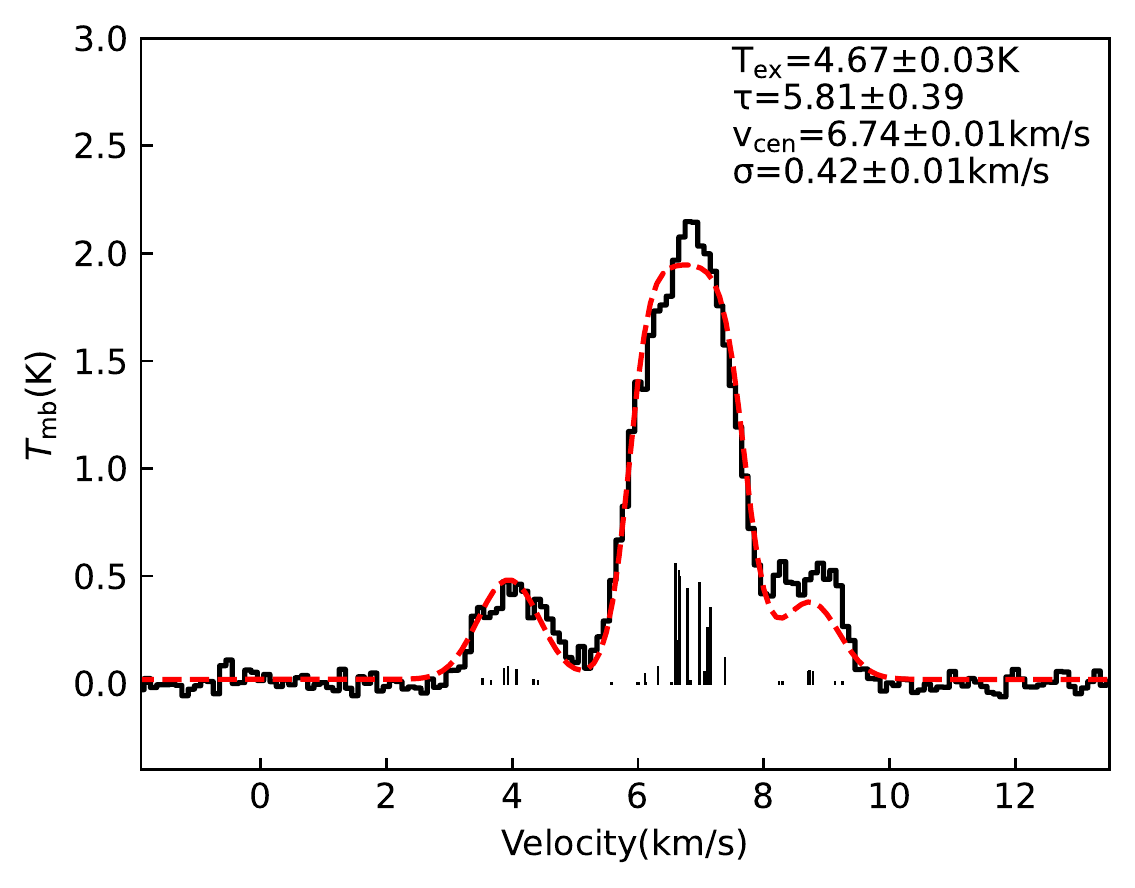}
\caption{\ntwohp(3–2) spectrum of CR1c1 shown in black. The relative intensities of hyperfine components are shown underneath the spectrum, also in this velocity frame.} The dashed red line shows the best-fit spectrum. The returned parameters from fitting (excitation temperature, opacity, centroid velocity, velocity dispersion) are displayed at the top right corner. The opacity listed is the total opacity by summing up the opacities of all the hyperfine components. The peak opacity in the best-fit case is slightly smaller ($\tau_{\rm peak}$ = 4.64).
\label{fig:fit}
\end{figure}

Under the assumption of constant excitation temperature among the hyperfine components of \ntwohp(3-2), the brightness temperature can be represented as:

\begin{equation}
    T_B(v)=[J(T_{\rm ex})-J(T_{\rm bg})][1-{\rm exp}(-\tau(v))]
\end{equation}
where $J(T)\equiv\frac{h\nu/k}{{\rm exp}(h\nu/[kT])-1}$ and $T_{\rm bg}$ = 2.73~K and the optical depths of the multiplets are

\begin{equation}
    \tau(v)=\tau_{\rm tot}\sum_{i}R_i {\rm exp}\left[ - \frac{(v-v_i-v_{\rm sys})^2}{2 \sigma^2}  \right],
\end{equation}
where $R_i$ and $v_i$ are the relative intensity and velocity for the $i$th hyperfine component, respectively. \autoref{fig:fit} shows the best-fit for the \ntwohp(3-2) spectrum of CR1c1 and the returned best-fit parameters are displayed on the top right corner.

\section{Infrared counterparts of the cores in Vela C CR1}
\label{sec:appB}

\begin{figure}[ht!]
\epsscale{0.93}\plotone{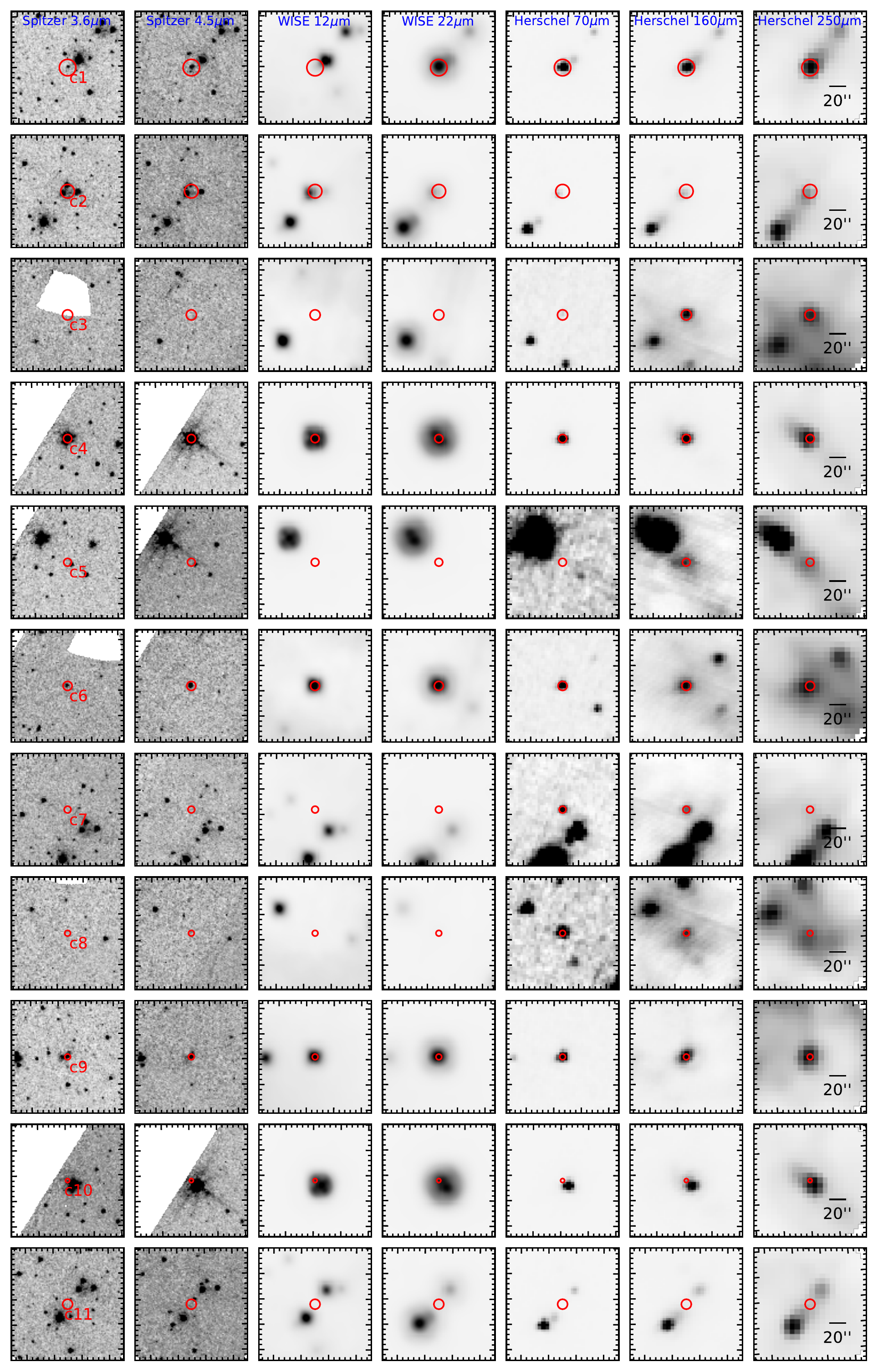}
\caption{Infrared counterparts of the cores in the CR1 region. From left to right we present the images of {\it Spitzer} 3.6/5.8~$\mu$m, {\it WISE} 12/22~$\mu$m and {\it Herschel} 70/160/250~$\mu$m, and from top to bottom the images are centered on the position from CR1c1 to CR1c11. The red circles indicate the position and size measured from the ALMA 1.3~mm continuum.}
\label{fig:ir}
\end{figure} 

\begin{figure}[ht!]
\epsscale{1.0}\plotone{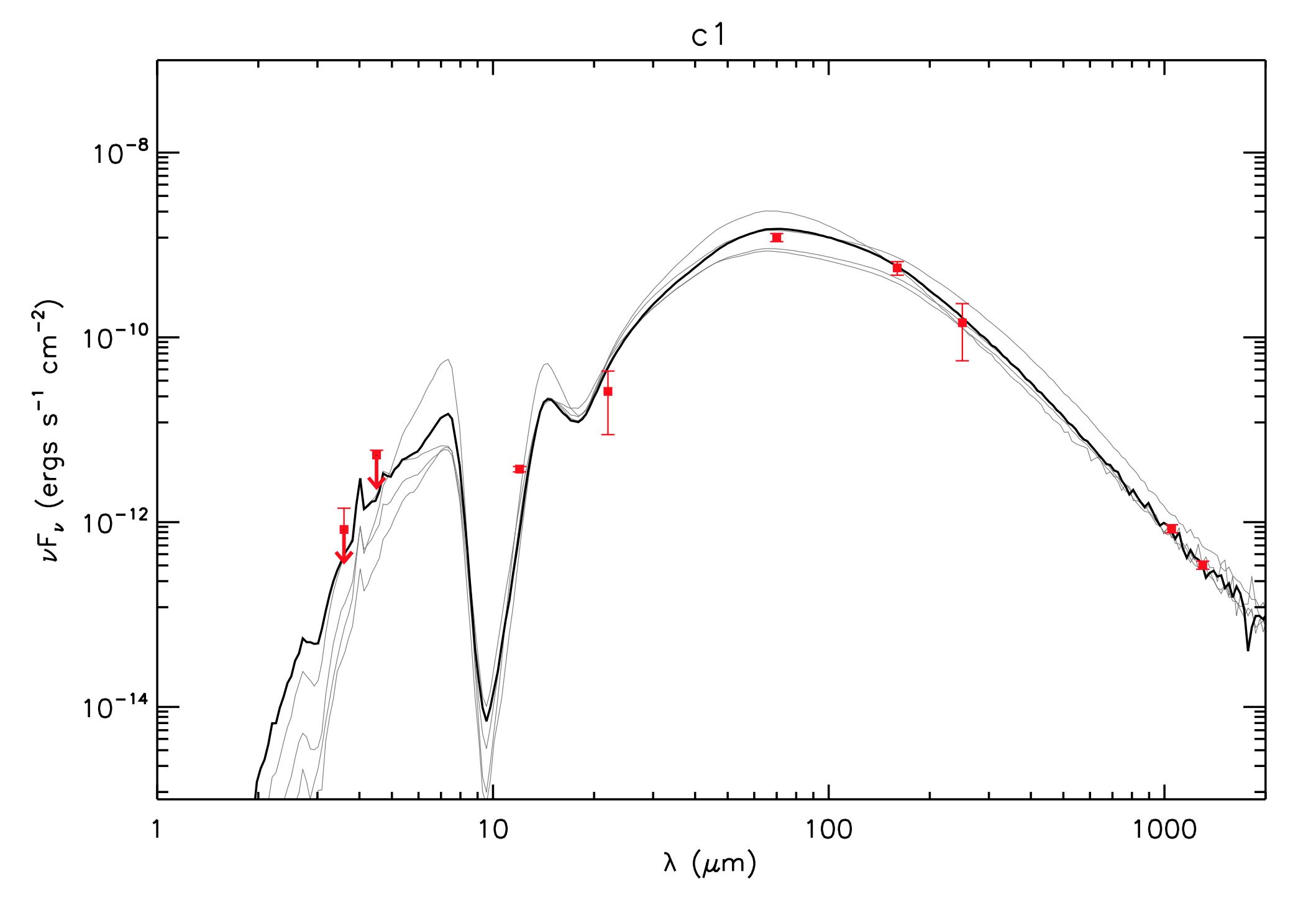}
\caption{Protostellar model fitting to the fixed aperture, background-subtracted SED of CR1c1 using the \citet{Zhang18} model grid. The best-fit model is shown with a solid black line and the next four best models are shown with solid gray lines. Flux values are those from \autoref{table:phot}. The fluxes in 1.05~mm and 1.3~mm are measured with the same background subtraction method, with values of 0.298~Jy and 0.147~Jy, respectively.}
\label{fig:c1_sed}
\end{figure}

In order to determine the evolutionary stages of dense cores, we searched for infrared counterparts from near-IR to far-IR bands. We retrieved the datasets in the archive including {\it Spitzer} 3.5, 4.5~$\mu$m, {\it WISE} 12, 22~$\mu$m and {\it Herschel} 70, 160, 250, 350, 500~$\mu$m maps. For each core we checked the infrared detection at the location and boundary defined by the ALMA 1.3~mm continuum to determine if there is a counterpart. \autoref{fig:ir} presents a zoom-in view of the infrared images for each core from 3.6~$\mu$m to 250~$\mu$m. The detection status is summarized in \autoref{table:phot}. Detections at $\lambda >$ 250~$\mu$m are not included here since the spatial resolution is poor and hence it is in general difficult to disentangle the dense core from the surrounding material.

As can be seen, CR1c11 is presumably in a very early evolutionary stage, since it is undetected in all infrared wavelengths. The detection of CR1c10 is confused by another adjacent bright source. CR1c5 is not seen at wavelengths up to 70~$\mu$m, but is detected weakly at 160~$\mu$m and 250~$\mu$m. CR1c3, CR1c7 and CR1c8, which are detected at 70~$\mu$m but not shorter wavelengths, are slightly more evolved and should have formed protostars. The remaining sources, i.e., CR1c1, CR1c2, CR1c4, CR1c6 and CR1c9, are clearly seen in all infrared wavelengths and are thus expected to be of the latest protostellar evolutionary stages among the sample.

 For infrared bands that show a counterpart, we also attempt aperture photometry, with the aperture radius fixed to be the equivalent radius of the size measured from 1.3~mm continuum. If the aperture radius is greater than half of the beam FWHM at the corresponding infrared band, then a flux measurement and associated uncertainty are reported. With this criterion only for CR1c1 are we able to measure the fluxes in all wavelengths. The aperture photometry is done following the method of \citet{Liu20}, i.e., we carry out a background subtraction using the median flux density in an annular region extending from one to two aperture radii, to remove general background and foreground contamination. The error bars are set to be the larger of either 10\% of the background-subtracted flux density or the value of the estimated background flux density. 

 To better constrain the physical parameters of dense core, we used \citet{Zhang18} radiative transfer models (ZT models hereafter) to fit the near-IR to millimeter SEDs towards CR1c1. The ZT model is a continuum radiative transfer model that describes the evolution of high- and intermediate-mass protostars with analytic and semi-analytic solutions based on the paradigm of the Turbulent Core model \citep[see][for more details]{Zhang18}. The main free parameters in this model are the initial mass of the core $M_c$, the mass surface density of the clump that the core is embedded in $\Sigma_{\rm cl}$, the protostellar mass $m_*$, as well as other parameters that characterize the observational setup, i.e., the viewing angle $i$, and the level of foreground extinction $A_V$. Properties of different components in a protostellar core, including the protostar, disk, infall envelope, outflow, and their evolution, are also derived self-consistently from given initial conditions. \autoref{fig:c1_sed} shows an example of the SED fit for CR1c1, with the parameters for the best five fitted models reported in \autoref{table:sed}. The best fitted model indicates a source with a protostellar mass of 2~\msun{} accreting at a rate of 3$\times10^{-5}$ $\rm M_\odot \cdot~yr^{-1}$ inside a core with an initial mass of 10~\msun{} embedded in clumps with a mass surface density of 0.3 $\rm g\cdot cm^{-2}$. Note that the ZT models are designed for high-mass star formation and $M_c=10\:M_\odot$ is the minimum core mass explored. Thus there could be other viable protostellar properties, e.g., starting with lower $M_c$, that have not been explored here.

\begin{deluxetable*}{cccccccc}
\tabletypesize{\scriptsize}
\caption{Photometry of cores in Vela C CR1 from 3.6 to 250~$\mu$m \tablenotemark{a}}
\label{table:phot}
\tablehead{
\colhead{Core} & & & & \colhead{Wavelength ($\mu$m) }& & &  \\ \hline
 & \colhead{3.6}& \colhead{5.8}  & \colhead{12} &\colhead{22} & \colhead{70}& \colhead{160} & \colhead{250} 
}
\startdata
c1	& 0.001 (0.0007)	&0.008 (0.001)	&0.015 (0.001)	&0.190 (0.126)	&28.189 (2.819)	&30.207 (5.041)	&11.992 (7.347)	\\
c2	& 0.017 (0.002)	&0.033 (0.003)	&0.018 (0.009)	&0.035 (0.031)	&2.049 (0.205)	&5.272 (1.721)	&Y	\\
c3	& N	&N	&N	&N	&0.057 (0.006)	&0.935 (0.646)	&Y	\\
c4	& 0.116 (0.012)	&0.791 (0.079)	&Y	&Y	&23.118 (3.764)	&Y	&Y	\\
c5	&N	&N	&N	&N	&N &Y	&Y	\\
c6	& 0.002 (0.0002)	&0.016 (0.002)	&Y	&Y	&0.499 (0.070)	&Y	&Y	\\
c7	& N	&N	&N	&N	&0.093 (0.027)	&Y	&Y	\\
c8	& N	&N &N	&N	&Y	&Y	&Y	\\
c9	& 0.003 (0.0003)	&0.013 (0.001)	&Y	&Y	&Y	&Y	&Y	\\
c10	& N	&N	&N	&N	&N	&N	&N\\
c11	&  N	&N	&N	&N	&N	&N	&N\\
\enddata
\tablenotetext{a}{The fluxes are in unit of Jy. Cores with and without an counterpart at corresponding wavelengths are indicated with 'Y' and 'N', respectively. If an infrared counterpart is detected and the core has a radius greater than half of the FWHM beam at corresponding infrared wavelength, we perform an aperture photometry (see text for more details) and the measured fluxes and uncertainties are shown (instead of 'Y').}
\end{deluxetable*}

\startlongtable
\begin{deluxetable*}{cccccccccccccc}
\tabletypesize{\scriptsize}
\renewcommand{\arraystretch}{1.0}
\tablecaption{Parameters of the best five fitted models for CR1c1\label{table:sed}}
\tablehead{
\colhead{Source} & \colhead{$\chi^2$}  & \colhead{$M_c$} & \colhead{$\Sigma_{\rm cl}$}& \colhead{$R_{\rm core}$} & \colhead{$m_*$} & \colhead{$\theta_{\rm view}$} & \colhead{$A_V$} &  \colhead{$\theta_{\rm w,esc}$} & \colhead{$m_{\rm disk}$} & \colhead{$r_{\rm disk}$} &  \colhead{$ {\dot{m}_{\rm disk}}$} & \colhead{$L_{\rm bol,iso}$} & \colhead{$L_{\rm bol}$} \\
\colhead{} & \colhead{}  & \colhead{$M_\odot$} & \colhead{$\rm g \cdot cm^{-2}$} & \colhead{$\rm pc$} & \colhead{$M_\odot$} & \colhead{\arcdeg} & \colhead{mag} &  \colhead{\arcdeg} & \colhead{$M_\odot$} & \colhead{(AU)} &  \colhead{$M_\odot/yr$} & \colhead{$L_\odot$} & \colhead{$L_\odot$}
}
\startdata
CR1c1 & 0.68 & 10 & 0.3 & 0.04 & 2.0 & 62 & 74.2 & 43 & 0.7 & 63 & 3.0 $\times10^{-5}$ & 2.8 $\times10^2$ & 0.7 $\times10^2$ \\
$R_{\rm ap}$ = 9\farcs{5} & 1.80 & 10 & 0.1 & 0.07 & 0.5 & 71 & 79.6 & 20 & 0.2 & 37 & 0.8 $\times10^{-5}$ & 0.8 $\times10^2$ & 0.5 $\times10^2$ \\
(0.04~pc) & 2.46 & 10 & 0.1 & 0.07 & 1.0 & 68 & 82.2 & 31 & 0.3 & 63 & 1.0 $\times10^{-5}$ & 1.1 $\times10^2$ & 0.5 $\times10^2$ \\
 & 5.07 & 20 & 0.1 & 0.10 & 0.5 & 86 & 72.9 & 13 & 0.2 & 32 & 1.0 $\times10^{-5}$ & 0.9 $\times10^2$ & 0.7 $\times10^2$ \\
 & 6.27 & 10 & 1.0 & 0.02 & 4.0 & 65 & 184.5 & 59 & 1.3 & 76 & 7.7 $\times10^{-5}$ & 11.4 $\times10^2$ & 1.7 $\times10^2$ \\
\enddata
\end{deluxetable*}

\bigskip
\vspace{5mm}
\facilities{Atacama Large Millimiter/submillimeter Array (ALMA)}


\software{CASA \citep{McMullin07}, APLpy \citep{Robitaille12}, Astropy \citep{Astro13}}

\acknowledgements This paper makes use of the following ALMA data:
ADS/JAO.ALMA\#2018.1.00227.S. ALMA is a partnership of ESO
(representing its member states), NSF (USA) and NINS (Japan), together
with NRC (Canada), NSC and ASIAA (Taiwan), and KASI (Republic of
Korea), in cooperation with the Republic of Chile. The Joint ALMA
Observatory is operated by ESO, AUI/NRAO, and NAOJ. The National Radio
Astronomy Observatory is a facility of the National Science Foundation
operated under cooperative agreement by Associated Universities, Inc.
Support for this work was provided by the NSF through the Grote Reber Fellowship Program (to Y.C.) administered by Associated Universities, Inc./National Radio Astronomy Observatory.
J.C.T. acknowledges support from NSF grant AST1411527, VR grant 2017-04522 and ERC project 788829 - MSTAR.





\bibliography{refer}
\end{document}